\RequirePackage[colorlinks]{hyperref}

\documentclass[draft]{agujournal2019}
\usepackage{url} 
\usepackage{lineno}
\usepackage{soul}
\usepackage{amsmath} 
\usepackage{comment}
\usepackage{booktabs}
\usepackage{amssymb}
\usepackage{enumerate}

\newcommand{\possessivecite}[1]{\citeauthor{#1}'s \citeyear{#1}}
\usepackage{tikz,pgfplots}
\usepgfplotslibrary{groupplots}
\usepackage{pgfplotstable}
\usepgfplotslibrary{external}
\usetikzlibrary{positioning,arrows.meta,shapes}
\pgfplotsset{compat=1.15}
\usetikzlibrary{intersections, pgfplots.fillbetween}
\usetikzlibrary{patterns}
\usetikzlibrary{quotes,arrows.meta}

\hypersetup{
    colorlinks = true,
    urlcolor   = black,
    citecolor  = black,
    linkcolor= black,
}

\draftfalse

\journalname{preprint}

\begin{document}

%
%


\title{Surface detachment and bed entrainment of fluvial plastics}

%
%




\authors{Matthias Kramer\affil{1}}


\affiliation{1}{UNSW Canberra, School of Engineering and Technology, Canberra, ACT 2610, Australia}




\correspondingauthor{Matthias Kramer}{m.kramer@unsw.edu.au}



\begin{keypoints}
\item Detachment velocities demonstrate that the free-surface acts as sink for positively buoyant plastics
\item Novel physical basis allows to reconcile the incipient motion of plastics with the classical Shields diagram
\end{keypoints}

%
%

%
%


\begin{abstract}
Over the last decade, fluvial plastics have been identified as major threat to aquatic environments and human health. In order to develop adequate mitigation strategies for plastic pollution, a fundamental process understanding of riverine plastic transport is of significant importance. In this context, the implementation of research findings into numerical simulation environments is anticipated to enhance modelling capabilities and to support a rigorous decision making. Recent experimental research has focused on the incipient motion of plastic particles, as well as on the effects of surface tension on plastic concentration profiles. While these investigations have advanced the state-of-the-art knowledge, current literature still displays a lack of basic insights into layer-specific plastic transport physics. In this study, first principles are applied to advance knowledge on free-surface detachment and bed entrainment of fluvial plastics. A novel relationship for the critical surface detachment velocity is derived, followed by the development of a framework that allows to relate plastic Shields parameters to those of natural sediments. Overall, it is anticipated that these developments will trigger new research within the plastics community, and it is hoped that present findings will be implemented into Lagrangian particle tracking software.  
\end{abstract}

\section{Introduction}
In recent years, the mechanics of fluvial plastic transport have been subject to more detailed investigations, and it was shown that different layers of plastic transport can be distinguished, including bed layer, suspended layer, and surface layer. This distinction is important, as these layers are governed by different flow physics. For example, plastic particles in the bed layer are interacting with sediment particles \cite{LOFTY2023120329}, suspended plastics are subject to water drag and lift, while surfaced plastics are additionally influenced by air drag and surface tension \cite{CHUBARENKO2016105,VALERO2022119078}.

Of the three mentioned layers and associated transport modes, suspended load transport and bed load transport constitute transport processes that have been studied extensively in sediment research, and as such, sediment research can provide valuable insights for plastic research \cite{WALDSCHLAGER2022104021}. One prominent example of this knowledge transfer is the adaptation of the well-known Rouse equation to positively buoyant plastics, as first discussed in \citeA{cowger2021concentration}. However, it is noted that there are important differences between natural sediments and plastics in suspended load  transport, which are primarily caused by varying shapes and materials. \citeA{Melk_2020} reviewed drag coefficients of plastics in suspension, concluding that the approaches of \citeA{bagh_2016} and \citeA{Dio_2018} are currently the most accurate to predict shape-dependent drag coefficients and terminal velocities of common plastics, such as disks, ellipsoids, cylinders, fibres, etc. Further, several effects on suspended particle drag and lift, including effects of turbulence, secondary motion, and hindered settling, warrant additional research. 

In bed load transport, the onset of motion of plastics is of fundamental importance. This phenomenon was first investigated by \citeA{Waldschlaeger2019} for plastics on different sediment bed configurations, who related the critical plastic Shields parameter ($\theta_{\text{cr},p}$) to the critical sediment Shields parameter ($\theta_{\text{cr},s}$) as follows
\begin{equation}
\frac{\theta_{\text{cr},p}}{\theta_{\text{cr},s}} = c_1 \underbrace{\left(\frac{D_p}{d_{50}}\right)^{c_2}}_{\text{\parbox{2cm}{\centering hiding-exposure \\[-4pt] function}}},
\label{EqWald}
\end{equation}
where $D_p$ is the representative plastic diameter, $d_{50}$ is the median grain size of the sediment bed, and the two empirical parameters were determined as $c_1 = 0.5588$ and $c_2 = -0.503$, the latter controlling the strength of the hiding-exposure effect. \citeA{Goral2023} interpreted the parameter $c_1$ as the ratio of static friction coefficients between plastic and sediment, and additionally modified 
the exponent of the so-called hiding-exposure function, originally introduced for sediments by \citeA{WILCOCK1988}. While Eq. (\ref{EqWald}), as well as modifications thereof, are useful for a first assessment of the incipient motion of plastic particles on sediment beds, their simplicity somewhat disguises the complexity of underlying physical processes, which is because different effects are lumped into the hiding-exposure function. In this context, it is evident that future in-depth studies of plastic re-suspension are required. According to \citeA{ROHAIS2024104822}, these studies should encompass a wider range of (micro-) plastic parameters and explore different definitions of the onset of motion, similar to those presented in \citeA{Yu2022,Yu2023}, aiming to derive general expressions for plastic re-suspension behaviour. Revising the literature on plastic re-suspension to date, it becomes clear that a better understanding of underlying physical processes is required, and that the applicability of sediment re-suspension models needs to be critically assessed. 

Considerably less research efforts have been devoted to particle transport processes of the free-surface layer. \citeA{CHUBARENKO2016105} were the first to consider air drag acting on floating plastics, further deriving an expression for the relative submergence of surfaced spheres. However, \citeA{CHUBARENKO2016105} did not consider surface tension forces in their analysis, implying that some of their presented equations must be revised. \citeA{VALERO2022119078} established the key importance of surface tension forces in surface load transport, demonstrating that surface tension effects on the concentration profile can be as intense as buoyancy, and that surface tension bias can lead to a drastic underestimation of total transported plastic. Despite these advances, the common understanding of the fundamental mechanics of plastic surface load transport remains limited, and further insights are required to improve theoretical and numerical modelling capabilities.

This study aims to establish some important underpinning foundations of particle transport mechanics in the bed and the free-surface layer, comprising free-surface detachment ($\S$ \ref{secFreeSurf}) and bed entrainment ($\S$ \ref{methods}) of fluvial plastics. In $\S$ \ref{secFreeSurf}, linear momentum conservation is applied to a floating plastic particle, enabling the derivation of a novel formulation for the particle floating velocity ($\S$ \ref{FloatingVel}) and the critical detachment velocity ($\S$ \ref{DetachmentVel}). These derivations are followed by a preliminary assessment, demonstrating that the free-surface acts as sink for positively buoyant microplastics. In $\S$ \ref{secShields}, the focus is set on the incipient motion of plastic particles, and a general formulation for the plastic Shields parameter is presented. As there is only limited experimental data available from literature, a framework that relates the plastic Shields parameter to that of natural sediment ($\S$ \ref{secInter}) is introduced, leading to the appearance of a shape factor ratio in our expanded equations. Subsequently, this framework is applied to the literature  data sets of \citeA{Waldschlaeger2019} and \citeA{Goral2023} in $\S$ \ref{secReanalysis}, and it is demonstrated mathematically that irregular shapes can lead to a reduced or increased mobility of plastics when compared to natural sediment.

\section{Free-surface detachment}
\label{secFreeSurf}

\subsection{Floating velocity}
\label{FloatingVel}
\begin{figure}[h!]
\begin{center}
\includegraphics[angle=0]{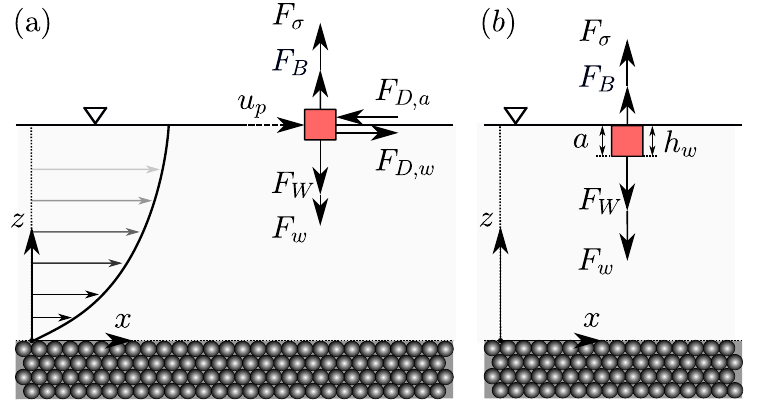}
\end{center}
\caption{Floating plastic in an open-channel flow; $x =$ streamwise coordinate, $z =$ vertical coordinate: (\textit{a}) forces acting on a floating plastic; (\textit{b}) detachment condition for a cubical particle with side length $a$ and submerged depth $h_w$}
\label{fig1}
\end{figure}

In the following, an expression for the velocity of a floating plastic particle at the free surface is derived. Let us consider a plastic particle with sphere-volume equivalent diameter $D_p = \sqrt[3]{6\mathcal{V}_p/\pi}$, volume $\mathcal{V}_p$, and density $\rho_p$, which is floating at the free-surface of an open-channel (Fig. \ref{fig1}\textit{a}). The particle moves with the velocity $u_p$ and is subject to water drag $F_{D,w}$, air drag $F_{D,a}$, buoyancy $F_B$, surface tension $F_\sigma$, weight force $F_W$, as well as vertical surface stresses and pressures that are exerted over the particle's surface by the water phase, summarised into the force $F_w$.

Implying that the acceleration of the floating particle is negligible, the acting forces are evaluated in streamwise direction, as shown in Fig. \ref{fig1}\textit{a}, further assuming no external wind forcing, i.e., \textcolor{black}{$u_a = 0$}, with $u_a$ being the air velocity, yielding
\begin{equation}
\underbrace{\frac{1}{2} \, \rho_w \, C_{D}  \, A_{\text{proj},w} \, \vert u_\text{fs} - u_p  \vert \,  (u_\text{fs} - u_p)}_{F_{D,w}}  = \underbrace{\frac{1}{2} \, \rho_a C_{D}  \, A_{\text{proj},a} \, u_p^2}_{F_{D,a}},
\label{forcebalancehorizontal}
\end{equation}
where $\rho_w$ is the water density, $C_{D}$ is the drag coefficient, $A_{\text{proj},w}$ is the submerged projected surface area of the plastic particle, $u_\text{fs}$ is the free-surface velocity, i.e., $u_\text{fs} = u(z=H)$, with $u$ being the streamwise velocity, $H$ the water depth, and $z$ the vertical coordinate.  Further, $A_{\text{proj},a}$ is the projected surface area of the plastic particle above the water surface, and $\rho_a$ is the density of air. In a first approximation, $C_D$ is assumed to be identical for the near-surface water layer and the air superlayer. Simplification leads to
\begin{equation}
\frac{\left(u_\text{fs} - u_p \right)^2}{u_p^2}= \frac{\rho_a}{\rho_w} \frac{A_{\text{proj},a}}{A_{\text{proj},w}},
\end{equation}
yielding an expression for the particle velocity without find forcing
\begin{equation}
u_p = \frac{u_\text{fs}}{\left(1  + \sqrt{\frac{\rho_a}{\rho_w} \frac{A_{\text{proj},a}}{A_{\text{proj},w}}} \right)}.
\label{eqparticlevelocity}
\end{equation}

\textcolor{black}{It is noted that the assumption $u_a = 0$ neglects the velocity of the air dragged by the water phase. If air was to move with the water surface velocity, i.e., $u_a = u_\text{fs}$, one would obtain $u_p = u_\text{fs}$. Given the unknown velocity distribution of the air superlayer, it is opted to keep the more general expression [Eq. (\ref{eqparticlevelocity})], noting that $u_p \lessapprox u_\text{fs}$, which is due to the large density difference between air and water, i.e., $\rho_a \ll \rho_w$.} 

Next, Newton's second law is applied to a floating plastic in vertical direction, where a distinction is made between body forces, i.e., gravity, and surface forces, including pressures and surface stresses. This application is detailed in \ref{App:vertical} and leads to the following expression for the vertical force balance (Fig. \ref{fig1}\textit{a})
\begin{equation}
\underbrace{\vphantom{\frac{1}{2}}\rho_w g \mathcal{V}_{p,w}}_{F_B} + \underbrace{\vphantom{\frac{1}{2}}L_\sigma \, \sigma \, \sin{(\Omega)}}_{F_\sigma} =  \underbrace{\vphantom{\frac{1}{2}}\rho_p g\mathcal{V}_p}_{F_W} + \underbrace{\frac{1}{2} \rho_w C_{D,{F_w}} A_{\text{proj},F_w} \, w'^2_\text{rms}}_{F_w},
\label{eqparticlevelocity1}
\end{equation}
where $F_B$ is the buoyancy force, $F_\sigma$ is the surface tension force, $F_W$ is the weight force, $F_w$ is the vertical force due to non-hydrostatic pressures and surface stresses, $g$ is the gravitational acceleration, $\mathcal{V}_{p,w}$ and $\mathcal{V}_p$ are the submerged particle volume and the total particle volume, respectively, $L_\sigma$ is the interfacial contact length, $\sigma$ is the surface tension,  $\Omega$ the contact angle, $\sin{(\Omega)}$ accounts for the vertical projection of the surface tension force, $w'_\text{rms}$ is the root-mean-square of vertical velocity flucuations, $C_{D,{F_w}}$ is a coefficient, and $A_{\text{proj},F_w}$ is the projected area for $F_w$. 

Subsequently, $w'_\text{rms}$ is characterised using a semi-empirical relationship for open-channel flows from \citeA{nezu1993turbulence} \begin{equation}
\frac{w'_\text{rms}}{u_*} = 1.27 \, \exp{\left(-\frac{z}{H}\right)},  
\label{TKE}
\end{equation}
where $u_*$ is the shear velocity. Next, we recall the  classical log-law for rough surfaces \cite{Dey2019} 
\begin{equation}
\frac{u_\text{fs}}{u_*} =\frac{1}{\kappa} \ln \left( \frac{z}{z_0} \right),  
\label{loglaw}
\end{equation}
where $z_0$ is the hydraulic roughness. Combining 
Eq. (\ref{eqparticlevelocity}) with Eq. (\ref{TKE}) and Eq. (\ref{loglaw}), the latter two evaluated at \textcolor{black}{at $z = H$}, leads to
\begin{equation}
w'_\text{rms}(H) = \frac{0.467 \, u_\text{fs} \, \kappa }{ \ln \left( \frac{H}{z_0} \right)} = \frac{0.467 \, u_p \, \kappa }{ \ln \left( \frac{H}{z_0} \right)} \left(1  + \sqrt{\frac{\rho_a}{\rho_w} \frac{A_{\text{proj},a}}{A_{\text{proj},w}}} \right).
\label{TKE2}
\end{equation}

Rearrangement and combination of Eq. (\ref{eqparticlevelocity1}) with Eq. (\ref{TKE2}) yields another expression for the velocity of the floating plastic
\begin{equation}
u_{p}   = \frac{\ln \left( \frac{H}{z_0} \right)}{ 0.467 \, \kappa } \, \frac{\sqrt{\frac{\rho_w g \mathcal{V}_{p,w}   - \rho_p g \mathcal{V}_{p} + L_\sigma \sigma \sin{(\Omega)}}{\frac{1}{2} \rho_w C_{D,{F_w}} A_{\text{proj},F_w} }}}{ 1  + \sqrt{\frac{\rho_a}{\rho_w} \frac{A_{\text{proj},a}}{A_{\text{proj},w}}}}.
\label{TKE3}
\end{equation}
\textcolor{black}{Although Eq. (\ref{TKE3}) was derived through application of Newton's second law in vertical direction, it represents an expression for the streamwise floating velocity of a plastic particle, yielding the same results as Eq. (\ref{eqparticlevelocity}). It is acknowledged that the relative areas $A_{\text{proj},a}/A_{\text{proj},w}$, as well as the forces $F_B$, $F_\sigma$, $F_W$, and $F_w$ are required for a computation of $u_p$.}

A dimensionless form can be obtained by dividing Eq. (\ref{TKE3}) with $\sqrt{\left( \frac{\vert \rho_w  - \rho_p \vert}{\rho_w} \right)g  D_p  }$, further introducing a shape factor $\beta_p = (A_{\text{proj},F_w} D_p) / \mathcal{V}_p$, as well as a modified plastic-based E\"otv\"os (or Bond) number $\Gamma$, representing a combination of weight, buoyancy, and surface tension forces
\begin{equation}
\Gamma = \frac{F_B - F_W + F_\sigma}{ \vert F_{B,\text{max} } -  F_W \vert} =   \frac{\rho_w  g \mathcal{V}_{p,w} - \rho_p g \mathcal{V}_{p}  + L_\sigma \sigma \sin{(\Omega)}}{ g \mathcal{V}_p (\vert\rho_w - \rho_p \vert)},
\label{eqGamma}
\end{equation}
where absolute values $\vert \rho_w  - \rho_p \vert$ are introduced to account for positively and negatively buoyant plastic particles, and $F_{B,\text{max}}$ stands for the maximum buoyancy force. Rearrangement yields a concise expression for the dimensionless particle velocity $\Theta_{p}$
\begin{equation}
\Theta_p = \frac{u_{p}}{ \sqrt{\left( \frac{ \vert \rho_w  - \rho_p \vert}{\rho_w} \right)g  D_p  }}   = \frac{\ln \left( \frac{H}{z_0} \right)}{ 0.467 \, \kappa } \, \, \frac{\sqrt{ \frac{2 \, \Gamma}{\beta_p C_{D,{F_w}}}}  }{1  + \sqrt{\frac{\rho_a}{\rho_w} \frac{A_{\text{proj},a}}{A_{\text{proj},w}}}}.
\label{dimensionless}
\end{equation}

\subsection{Detachment velocity}
\label{DetachmentVel}
In this section, the position of a plastic particle as it is detaching from the interface is considered, allowing us to develop dimensional and dimensionless formulations for the surface detachment velocity, which can be regarded as the counterpart to well-known threshold velocity formulations for bed entrainment \cite[Chapter 4.3]{dey2014fluvial}. Herein, surface detachment formulations are presented for both, an arbitrarily shaped particle and a cubical particle, while spherical particles are discussed in more detail in $\S$ \ref{AppA}. Note that detachment conditions are indicated with the subscript ``cr'', and they apply to the parameters $u_p$, $\mathcal{V}_{p,w}$, $L_\sigma$, and $A_{\text{proj},a}/ A_{\text{proj},w}$. Inserting these conditions into Eq. (\ref{TKE3}) yields a general formulation for the critical particle velocity $u_{\text{cr},p}$ at surface detachment 
\begin{equation}
\label{eqverticalforce0}
u_{\text{cr},p} = 
\frac{\ln \left( \frac{H}{z_0} \right)}{ 0.467 \, \kappa } \, \, \frac{\sqrt{\frac{\rho_w g (\mathcal{V}_{p,w})_\text{cr} - \rho_p g \mathcal{V}_p  + L_{\text{cr},\sigma} \sigma \sin{(\Omega)}}{\frac{1}{2} \rho_w C_{D,{F_w}} A_{\text{proj},F_w} }}}{ 1  + \sqrt{\frac{\rho_a}{\rho_w} \left(\frac{A_{\text{proj},a}}{A_{\text{proj},w}} \right)_\text{cr} }}.
\end{equation}
which holds for all close-shaped particles. It is emphasized that Eq. (\ref{eqverticalforce0}) resembles a novel formulation for the detachment of floating plastics from the free-surface of an open-channel flow, where the critical velocity $u_{\text{cr},p}$ decreases with increasing bed roughness, particle weight, and hydrodynamic lift, while it increases with increasing water depth, surface tension, and buoyancy.

Next, let us define the detachment condition for a floating cubical particle. In view of Fig. \ref{fig1}\textit{b}, it becomes clear that detachment happens as the cubical particle is almost fully submerged, but still exposed to surface tension forces. In order to express this condition mathematically, the relative submergence of the particle is defined as $h_w/a$, where $a$ is the side length, and $h_w$ is the submerged depth (Fig. \ref{fig1}\textit{b}). At detachment, the relative submergence $(h_w/a)_\text{cr} \approx 1$, which implies that $\left(A_{\text{proj},a}/A_{\text{proj},w} \right)_\text{cr} \approx 0$. Inserting these conditions into Eq. (\ref{eqverticalforce0}), one can simplify
\begin{equation}
\left(u_{\text{cr},p}\right)_\text{cube} =
\frac{\ln \left( \frac{H}{z_0} \right)}{0.467 \, \kappa } \, \,  \sqrt{\frac{g a^2 (\rho_w - \rho_p) + 4 \, \sigma \sin{(\Omega)}}{\frac{1}{2} \rho_w C_{D,{F_w}} a}},
\label{eqverticalforce}
\end{equation}
where $\mathcal{V}_{p} = a^3$, $(\mathcal{V}_{p,w})_\text{cr} = a^3$, $L_{\text{cr},\sigma} = 4a$, and $A_{\text{proj},F_w} = a^2$ have been used. 

In a next step, a generalized dimensionless version of the critical detachment velocity is obtained by inserting the detachment conditions 
for the parameters $\Theta_p, u_p$, $\Gamma$, and $A_{\text{proj},a}/A_{\text{proj},w}$
into Eq. (\ref{dimensionless}), yielding 
\begin{equation}
\Theta_{\text{cr},p} =  \frac{u_{\text{cr},p}}{ \sqrt{\left( \frac{ \vert \rho_w  - \rho_p \vert}{\rho_w} \right)g  D_p  }}   = \frac{\ln \left( \frac{H}{z_0} \right)}{ 0.467 \, \kappa } \, \, \frac{\sqrt{ \frac{2 \, \Gamma_\text{cr}}{\beta_p C_{D,{F_w}}}}  }{1  + \sqrt{\frac{\rho_a}{\rho_w} \left(\frac{A_{\text{proj},a}}{A_{\text{proj},w}}\right)_\text{cr}}},
\label{dimensionless1}
\end{equation}
which can be regarded as the surface detachment counterpart of the well-known Shields parameter for bed entrainment. Importantly, detachment conditions need to be defined for each particle shape. For a cubical particle, Eq. (\ref{dimensionless1})
simplifies to
\begin{equation}
\left(\Theta_{\text{cr},p} \right)_\text{cube} = \frac{u_{\text{cr},p}}{ \sqrt{\left( \frac{ \vert \rho_w  - \rho_p \vert}{\rho_w} \right)g  D_p  }}    = \frac{\ln \left( \frac{H}{z_0} \right)}{ 0.467 \, \kappa } \sqrt{ \frac{2 \, \Gamma_\text{cr} }{ C_{D,{F_w}} \beta_p}},
\label{dimensionless2}
\end{equation}
where $\Gamma_\text{cr} = (\rho_w - \rho_p)/(\vert \rho_w - \rho_p \vert) + 4 \sigma \sin{(\Omega)}/g a^2 (\vert \rho_w - \rho_p \vert )$, as per Eq. (\ref{eqGamma}).

\begin{figure}[h!]
\begin{center}
\includegraphics[angle=0]{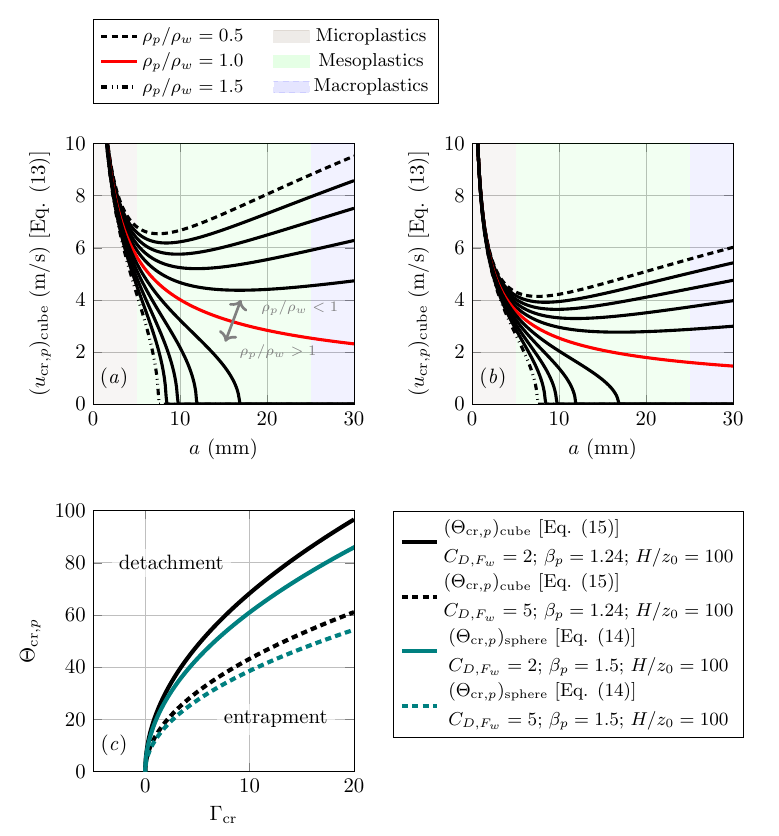}
\end{center}
\caption{Critical particle detachment velocity:  (\textit{a}) dimensional detachment velocity for a cubical plastic versus particle size for a coefficient $C_{D,{F_w}} = 2$; the simulated channel is characterised by $H/z_0 = 100$; (\textit{b}) same as (\textit{a}) but with $C_{D,{F_w}} = 5$; (\textit{c}) dimensionless critical detachment velocity $\Theta_\text{cr}$ versus inverse plastic-based E\"otv\"os number $\Gamma_\text{cr}$ at detachment; note that the shape factors $\beta_p = 1.24$ and 1.5 reflect a cubical and a spherical particle, respectively}
\label{fig2c}
\end{figure}

To provide a preliminary assessment of the derived formulations, Eq. (\ref{eqverticalforce}) is evaluated for a cubical particle floating at the free-surface of an open channel flow, the latter characterised by a ratio of water depth to hydraulic roughness $H/z_0 = 100$. To account for surface tension, an interfacial contact length $L_\sigma = 4a$, surface tension of $\sigma = 0.072$ N/m, and an interfacial contact angle $\Omega = 105^\circ$ are assumed, which is aligned with previously measured contact angles for plastics, for example 97.0$^\circ$ $\pm$ 0.8$^\circ$ \cite[PP at 25$^\circ$ C]{Vlaeva2012}, 96.5$^\circ$ $\pm$ 3.4$^\circ$ \cite[HDPE at 25$^\circ$ C]{deLuna2014}, between 93$^\circ$ and 105$^\circ$ \cite[PE]{ACCU2009PE}, and between 95$^\circ$ and 117$^\circ$ \cite[PP]{ACCU2009PP}. Taking the above into consideration, Eq. (\ref{eqverticalforce}) is solved for particle dimensions between $0 \leq a \leq 30$ mm, thereby accounting for micro- ($a < 5$ mm), meso- ($5 \leq a \leq 25$ mm), and macroplastics ($a > 25 $ mm). Note that in the absence of better knowledge, coefficients of $C_{D,{F_w}} = 2$ and $5$ have been selected, which requires experimental verification. Figures \ref{fig2c}\textit{a,b} show the results of this analysis, where $u_{\text{cr},p}$ is plotted against $a$ for eleven different density ratios, ranging from $0.5 \leq \rho_p/\rho_w \leq 1.5$, separated by increments of 0.1. Neutrally buoyant particles are indicated by the red lines, whereas positively and negatively buoyant particles are located above and below these solutions, respectively.  \textcolor{black}{In terms of physical implications for plastic surface load transport, it becomes clear that the free-surface acts as sink for positively buoyant plastics, which is further corroborated by the equivalent results for spherical particles ($\S$ \ref{AppA}). However, it is important to note that the detachment velocities shown in Fig. \ref{fig2c} do not provide information on whether particles do reach the free-surface. Depending on their settling velocity, negatively buoyant plastics may remain in the bed- and suspended layer, and it is anticipated that these particles are predominantly transported as bed load or suspended load.}

The evaluation of the dimensionless Eq. (\ref{dimensionless2}) for cubical particles is straightforward, and $(\Theta_{\text{cr},p})_\text{cube}$ is herein computed using $H/z_0 = 100$, $\kappa =0.41$, $(\beta_p)_\text{cube} = \sqrt[3]{6/ \pi} = 1.24$,  further assuming two coefficients, $C_{D,{F_w}} = 2$ and 5. Figure \ref{fig2c}\textit{c} shows these results, where dimensionless particle velocities $\Theta_{p} < \Theta_{\text{cr},p}$ and $\Theta_{p} \geq \Theta_{\text{cr},p}$ indicate surface entrapment and detachment, respectively. For spherical particles, the term \\ $\left(A_{\text{proj},a}/A_{\text{proj},w} \right)_\text{cr} > 0$, which requires an evaluation of Eq. (\ref{dimensionless1}). Such calculations are a bit more involved and are presented in more detail in $\S$ \ref{AppA}, while the results have been added to Fig. \ref{fig2c}\textit{c}, demonstrating detachment velocities for spherical particles are smaller than for cubical particles, which is because $(L_{\text{cr},\sigma})_\text{sphere} < (L_{\text{cr},\sigma})_\text{cube}$, given that $(D_p)_\text{sphere}$ = $(D_p)_\text{cube}$. Overall, most of the parameters in the generalized equations for surface detachment, i.e., Eqns. (\ref{eqverticalforce0})  and (\ref{dimensionless1}), can be determined or estimated, however, the surface tension force and the coefficient $C_{D,{F_w}}$ require future detailed experimental investigations, which are however beyond the scope of the present work. Lastly, it is stressed that Fig. \ref{fig2c}\textit{c} reflects the surface detachment counterpart of the Shields diagram for bed entrainment, and that the novel formulations, i.e., Eqns. (\ref{eqverticalforce0}) or (\ref{dimensionless1}), can be implemented into Lagrangian particle tracking simulations with relative ease.

\section{Bed entrainment}
\label{methods}

\subsection{Plastic Shields parameter}
\label{secShields}
\begin{figure}[h!]
\begin{center}
\includegraphics[angle=0]{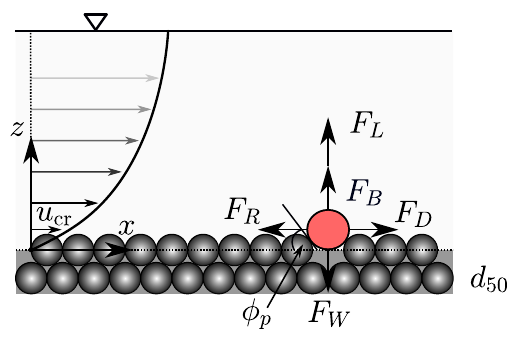}
\end{center}
\caption{Forces acting on a deposited plastic particle in an an open-channel flow; \textcolor{black}{the sediment bed is considered horizontal and sketched wit a uniform grain size $d_{50}$}}
\label{fig2}
\end{figure}

To derive an expression for the plastic Shields parameter, let us consider a plastic particle to be located on a sediment bed (Fig. \ref{fig2}), where the latter is characterised by its characteristic diameter $d_{50}$. The particle is subject to drag force $F_D$, resistance force $F_R$, weight force $F_W$, buoyancy $F_B$, and lift $F_L$. It is acknowledged that consideration of these forces corresponds to many derivations that exist for natural sediments, and the reader is referred to \citeA[Chapter 4]{dey2014fluvial} for an overview. As the flowrate  increases gradually, the plastic particle will eventually start moving. The corresponding velocity at particle level, which is adequate to initiate particle motion, is commonly referred to as critical velocity $u_{\text{cr}}$. In contrast to the critical particle velocity at the free-surface, introduced in $\S$ \ref{secFreeSurf}, the critical velocity $u_{\text{cr}}$ refers to the streamwise velocity of the fluid in vicinity of the channel bed. 

The application of Newton's second law to the plastic particle in horizontal direction  yields (Fig. \ref{fig2})
\begin{equation}
\underbrace{
 \frac{1}{2} \rho_w \, u_\text{cr}^2  C_{D,p}  A_\text{proj}}_{= F_D}  \, = \,   \underbrace{\vphantom{\frac{1}{2} \rho_w \, u_\text{cr,p}^2  C_{D,p}  A_\text{proj}}   \left( (\rho_p -\rho_w) g  \mathcal{V}_{p} - \frac{1}{2} \rho_w \, u_\text{cr}^2  C_{L,p} A_\text{proj} \right) \tan{(\phi_p)} }_{= F_R},
\label{eq1}
\end{equation}
where $C_{D,p}$ and $C_{L,p}$ are the plastic drag and lift coefficients, $A_\text{proj}$ is the projected area, $\tan{(\phi_p)}$ expresses the friction coefficient between the plastic and underlying sediment bed, and it is implied that the rate of change of particle momentum is zero.
\textcolor{black}{
While the sediment bed in Fig. \ref{fig2} is sketched with uniform grain size, it is important to note that Eq. (\ref{eq1}), specifically the formulation of the friction coefficient $\tan{(\phi_p)}$, can be used to represent entrainment on natural sediment beds, thereby taking plastic-sediment interaction (exposure and hiding effects), as well as different modes of entrainment, such as sliding and rolling, into consideration \cite{https://doi.org/10.1111/j.1365-3091.1990.tb00627.x,Wiberg1987,https://doi.org/10.1029/2023JF007162,doi:10.1061/TACEAT.0008054,https://doi.org/10.1111/j.1365-3091.1966.tb01897.x}}.

The hydrodynamic force on the sediment particle in Fig. \ref{fig2} is composed of hydrodynamic drag  and hydrodynamic lift, which are dependent on the particle Reynolds number $Re_p = (u_{\text{cr}} D_p)/\nu$. We note that the plastic particle is subject to a velocity gradient, and a strong dependence of drag and lift with respect to the shear rate can be expected. Further, the lift force at entrainment may comprise several components, including shear lift, Magnus lift, centrifugal lift, and turbulent lift, while it is stressed that there is no generic consensus on the lift force at particle entrainment \cite{DeyLift}. Nonetheless, the lift force is kept in the following, as it is regarded important for incipient motion. 

Rearranging and simplifying Eq. (\ref{eq1}) as per $\S$ \ref{ShieldsSediment} leads to a definition of the plastics Shields parameter $\theta_{\text{cr},p}$ 
\begin{equation}
\theta_{\text{cr},p}  = \frac{\tau_{\text{cr},p} }{ \left(\rho_p - \rho_w \right)  g D_p } = \frac{2}{\beta_p}   \underbrace{\frac{1}{\alpha_p^2}}_{=f(Re_{*}, \frac{D_p}{d_{50}})  } \underbrace{\frac{\tan{(\phi_p)}}{ C_{D,p}  + C_{L,p} \tan{(\phi_p)} }}_{=f(Re_{*}, \alpha_p, \frac{D_p}{d_{50}})},
\label{eq6}
\end{equation}
where $\tau_{\text{cr},p}$ is the critical shear stress at entrainment, the parameter $\alpha_p = u_{\text{cr}}/u_{*,\text{cr}}$ accounts for deviations between the critical velocity $u_{\text{cr}}$ and the shear velocity $u_{*,\text{cr}} = \sqrt{\tau_{\text{cr},p}/\rho_w}$, and the parameter $\beta_p = (A_\text{proj} D_p) / \mathcal{V}_p$ is a shape factor. We note that the term $1/\alpha_p^2$ is a function of $D_p/d_{50}$ and of the shear Reynolds number
$Re_{*} = (u_{*,\text{cr}} \, d_{50})/\nu$, whereas the combined drag-lift-friction term on the right hand side of Eq. (\ref{eq6}) is a function of $D_p/d_{50}$ and of the particle Reynolds number $Re_p$. Importantly, particle and shear Reynolds numbers are related to one another by 
\begin{equation}
\frac{Re_p}{Re_*} = \frac{u_{\text{cr}}}{u_{*,\text{cr}}} \frac{D_p}{d_{50}} = \alpha_p  \frac{D_p}{d_{50}}, 
\end{equation}
which allows us to write Eq. (\ref{eq6}) as function of $Re_{*}$, $\alpha_p$, and $D_p/d_{50}$ only. At this stage, it is important to recall that the incipient motion of plastic particles on a sediment layer is \textit{fully explained} by Eq. (\ref{eq6}). The unknowns are
the parameters $\alpha_p$ and $\beta_p$, the friction angle $\phi_p$, as well as plastic drag and lift coefficients, $C_{D,p}$ and $C_{L,p}$. 

\begin{figure}[h!]
\begin{center}
\includegraphics[angle=0]{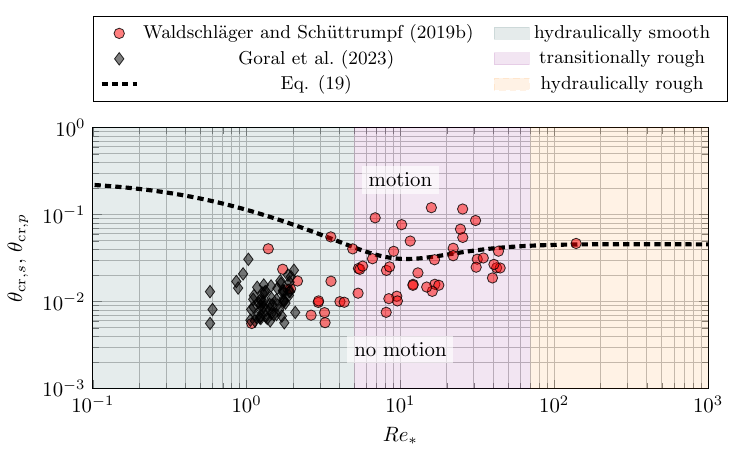}
\end{center}
\caption{Classical Shield diagram for sediment particles; plastic data are computed and plotted for the experiments of \citeA{Waldschlaeger2019} and \citeA{Goral2023}; the flow regimes are classified as hydraulically smooth flow for $Re_* \leq 5$, as hydraulically transitional flow for $5 < Re_* < 70$, and as  hydraulically rough flow for $Re_* \geq 70$}
\label{fig3}
\end{figure}

Figure \ref{fig3} shows the classical Shields diagram, where the Shields parameter is plotted against the shear Reynolds number. Here, microplastic data \cite{Waldschlaeger2019,Goral2023} have been included, and, for comparison, an analytical expression for the critical  Shields parameter of natural sediments $\theta_{\text{cr},s}$ \cite{Sui2021}
\begin{equation}
\theta_{\text{cr},s} = 0.165 (Re_{*} + 0.6)^{-0.8} + 0.045 \, \exp(-40 \, Re_{*} ^{-1.3}).  
\label{eqSui}
\end{equation}

It is seen that plastics have a higher mobility
when compared to natural sediment (Fig. \ref{fig3}), and that there is a large variability in the collected plastic data, which stems from different shapes and materials, but also from different sizes of the tested particles. If one was to establish a Shields curve for plastics as a complement to Eq. (\ref{eqSui}), some comprehensive testing, encompassing a wide range of plastic shapes, materials, and sizes would be required. It is also anticipated that there is not one general Shields curve for plastics, but rather different curves for different plastic shapes. 

\subsection{Plastic-sediment interrelation}
\label{secInter}
To advance knowledge on the incipient motion of plastic particles, it seems intuitive to seek a relationship between the plastic Shields parameter $\theta_{\text{cr},p}$ and the well-known Shields parameter for natural sediment $\theta_{\text{cr},s}$, which is presented in the following. Similar to the derivation of Eq. (\ref{eq6}), linear momentum conservation is applied to a sediment particle on a sediment bed, leading to an analytical equation for the sediment Shields parameter \cite{Wiberg1987,dey2014fluvial}
\begin{equation}
\theta_{\text{cr},s}  = \frac{\tau_{\text{cr},s} }{ \left(\rho_s - \rho_w \right)  g D_s } =  \frac{2 }{\beta_s} \underbrace{\frac{1}{ \alpha_s^2}}_{= f(Re_{*})} \, \, \underbrace{\frac{\tan{(\phi_s)}}{ C_{D,s}  + C_{L,s} \tan{(\phi_s)} }}_{= f (Re_{*}, \alpha_s)},
\label{eq5}
\end{equation}
where the subscript $s$ stands for sediment. We note that for natural sediments, $D_s \approx d_{50}$, which implies that the dependence on $D_s/d_{50}$ vanishes, i.e., $Re_s = \alpha_s Re_{*}$.  Dividing Eq. (\ref{eq6}) by Eq. (\ref{eq5}) yields two expressions that relate the plastic Shields parameter to the sediment Shields parameter
\begin{eqnarray}
\label{eq7a}
\frac{\theta_{\text{cr},p}}{\theta_{\text{cr},s}}  &=&   
\frac{\tau_{\text{cr},p}}{\tau_{\text{cr},s}}
\, \, 
\frac{\left( \rho_s - \rho_w \right) }{\left( \rho_p - \rho_w \right) } 
\, \, 
\frac{D_s}{D_p} \\
&=& 
\frac{\beta_s}{\beta_p} 
\, \, 
\frac{\alpha^2_s}{\alpha^2_p} 
\, \,  
\frac{C_{D,s}  + C_{L,s} \tan{(\phi_s)}}{C_{D,p} +  C_{L,p} \tan{(\phi_p)}} 
\, \, 
\frac{\tan{(\phi_p)}}{\tan{(\phi_s)}}. 
\label{eq7b}
\end{eqnarray}

It is important to note that Eqns. (\ref{eq7a}) and (\ref{eq7b}) are not only applicable to plastics. Rather, they can be used to relate any Shields parameter of a foreign particle $p$ to the Shields parameter of natural sediment, thereby representing a general and versatile framework. Note that there exists a variety of other expressions for the Shields parameter, some of them summarised in \citeA[Chapter 4]{dey2014fluvial}, which could alternatively be used to derive a relationship between the onset of motion of plastics and sediments.  

In the context of practical application of the proposed framework, it is important to give considerations to the flow situations that are being related to one another. For example, one might be interested in comparing the incipient motion of foreign particles with sediment particles assuming identical $Re_*$, which could further imply $D_s = d_{50}$, while the diameter of the foreign particle could be smaller or larger than the sediment particle, i.e., $D_p < D_s$ or $D_p > D_s$. To provide a second example, if the foreign particle is natural sediment with $D_p = D_s = d_{50}$, $\rho_p = \rho_s$, and other parameters being identical, the right hand sides of Eqns. (\ref{eq7a}) and (\ref{eq7b}) become unity, implying that the classical Shields diagram is recovered, i.e., $\theta_{\text{cr},p} = \theta_{\text{cr},s}$. Further, some of the involved terms, for example drag and lift coefficients of partially exposed plastics and sediments in the bed-boundary layer, are not well understood, which is one of the main reasons that the sediment Shields curve is of semi-empirical nature. Therefore, simplifications are required to proceed. \textcolor{black}{Resolving  Eqns. (\ref{eq7a}) and (\ref{eq7b}) for the critical shear stress ratio, it is obtained
\begin{equation}
\frac{\tau_{\text{cr},p}}{\tau_{\text{cr},s}}
= 
\frac{\left( \rho_p - \rho_w \right) }{\left( \rho_s - \rho_w \right) } 
\, \, 
\frac{D_p}{D_s}
\, \,
\frac{\beta_s}{\beta_p} 
\, \, 
\frac{\tan{(\phi_p)}}{\tan{(\phi_s)}} \, 
\underbrace{\frac{\alpha^2_s}{\alpha^2_p} 
\, \,  
\frac{C_{D,s}  + C_{L,s} \tan{(\phi_s)}}{C_{D,p} +  C_{L,p} \tan{(\phi_p)}} 
\, \, 
}_{=\left(\frac{D_p}{D_s}\right)^{c_2}}.
\label{Eq23}
\end{equation}}

\textcolor{black}{To simplify Eq. (\ref{Eq23}), the combined drag-lift-friction term on the right hand side is now replaced with a power law of the diameter ratio, also referred to as hiding-exposure function, yielding the following expression for the critical shear stress ratio 
\begin{equation}
\frac{\tau_{\text{cr},p}}{\tau_{\text{cr},s}}
= 
\frac{\left( \rho_p - \rho_w \right) }{\left( \rho_s - \rho_w \right) } 
\, \,
\frac{\tan{(\phi_p)}}{\tan{(\phi_s)}}
\, \,
\frac{\beta_s}{\beta_p} 
\, \, 
\left(\frac{D_p}{D_s}\right)^{(1 + c_2)}, 
\label{Eq24}
\end{equation}}
\textcolor{black}{where $c_2$ is the power law exponent. With regards to the simplification of the drag-lift-friction term, it can be argued that $\alpha_s^2/\alpha_p^2 \propto \left(D_p/D_s\right)^{c_4}$, with $c_4$ being an unknown exponent. It is however acknowledged that particle Reynolds-number effects on the drag- and lift-terms may not be completely captured. Further, it is worthwhile pointing out that Eq. (\ref{Eq24}) can be reconciled with \possessivecite{WILCOCK1988} empirical formula for natural sediments, i.e., $\tau_{\text{cr},p} / \tau_{\text{cr},s} = \left(D_p/D_s\right)^{c_3}$, by shifting the term $\tan{(\phi_p)}/ \tan{(\phi_s)}$ into the hiding-exposure function, further assuming that the foreign particle is a sediment particle, which implies $\rho_p = \rho_s$, and $\beta_p = \beta_s$.} \textcolor{black}{Returning to plastics, Eq. (\ref{Eq24}) is now substituted into Eq. (\ref{eq7a}), leading to the following expression for the plastic Shields parameter 
\begin{equation}
\theta_{\text{cr},p} = \theta_{\text{cr},s} \,
\frac{\beta_s}{\beta_p} 
\, \,
\frac{\tan{(\phi_p)}}{\tan{(\phi_s)}} 
\, \, 
\left(\frac{D_p}{D_s}\right)^{c_2}.
\label{eq13}
\end{equation}}

\textcolor{black}{
It is noted that Eq. (\ref{eq13}) constitutes an expansion of the approach suggested by \citeA[Eq. 4]{WALDSCHLAGER2019} and \citeA[Eq. 16]{Goral2023}, based on a rigorous application of linear momentum conservation. The same expression can also be obtained by only considering Eq. (\ref{eq7b}), however, the contemplation of Eq. (\ref{eq7a}) has produced an additional expression for the critical shears stress ratio, which was shown to be in alignment with an empirical formula for natural sediments, originally proposed by \citeA{WILCOCK1988}.}

\textcolor{black}{To summarize, the schematized physical basis for estimating the plastic Shields parameter, i.e., Eq. (\ref{eq13}), reveals that different physical effects have been lumped into a single hiding-exposure function. The presented derivation offers opportunities for future improvements, such that the individual effects of each parameter can be carefully examined and quantified. Herein, the expanded Eq. (\ref{eq13}) explicitly takes into account the particle shape by retaining the term $\beta_s/\beta_p$, which constitutes a theoretical and practical improvement over previous studies. Despite these advances, there is an imminent need for more fundamental testing on the mobilization of plastics, which should scrutinize how Eqns. (\ref{eq7a}) and  (\ref{eq7b}), or other versions thereof, can be simplified.}

\subsection{Data re-analysis}
\label{secReanalysis}
In this section, the expanded approach for estimating plastic Shields parameters [Eq. (\ref{eq13})] is applied to previous experimental data sets from literature. Note that the incipient motion of plastics and sediments is compared for identical $Re_*$, which is the most natural choice, as the calculation of particle Reynolds numbers would require some assumptions in the determination of $\alpha_p$ and $\alpha_s$. \textcolor{black}{Referring to Eq.  (\ref{eq13}), we seek an estimation of the unknown exponent $c_2$, while remaining parameters were available from previous studies on bed entrainment \cite{Waldschlaeger2019,Goral2023}, or from other information presented in this work. Parameters were determined as follows
\begin{enumerate}[(i)]
    \item Plastic Shields parameters $\theta_{\text{cr},p}$ were directly calculated from measured shear velocities as $\theta_{\text{cr},p} = u_{*,\text{cr}}/((\rho_p/\rho_w -1) g D_p)$ [Eq. (\ref{eqShieldsustar})]
    \item Sediment Shields parameters $\theta_{\text{cr},s}$ were estimated using Eq. (\ref{eqSui})
    \item Following \citeA{Goral2023}, friction terms were assumed for both experiments as $\tan{(\phi_p)}/\tan{(\phi_s)} = 0.5588$ and 0.55, respectively
    \item Sediment shape factors were considered as $\beta_s = 1.5$ (spherical), while plastic shape factors $\beta_p$ were computed using geometric particle specifications provided in Tab. \ref{table:conditions}. It is noted that the determination of $\beta_p$ for some tested shapes, such as fragments, fibers, etc., is not straightforward, and for those shapes, it was simply assumed $\beta_p/\beta_s = 1$, which allowed to retain these data.
\end{enumerate}}

\textcolor{black}{
Next, Eq. (\ref{eq13}) is multiplied with  $\frac{1}{\theta_{\text{cr},s}}\frac{\beta_p}{\beta_s}\frac{\tan{(\phi_s)}}{\tan{(\phi_p)}}$ to yield 
\begin{equation}
\frac{\theta_{\text{cr},p}} {\theta_{\text{cr},s}} \, \,
\frac{\beta_p}{\beta_s} \, \,     
\frac{\tan{(\phi_s)}}{\tan{(\phi_p)}}
 =
\left(\frac{D_p}{D_s}\right)^{c_2}
\label{eq2},
\end{equation} 
which can be solved explicitly for the power law exponent by taking the logarithm of both sides 
\begin{equation}
c_2 = \frac{\log \left(\frac{\theta_{\text{cr},p}}{\theta_{\text{cr},s}} \, \,
\frac{\beta_p}{\beta_s}  \, \,   
    \frac{\tan{(\phi_s)}}{\tan{(\phi_p)}}\right)}{ \log\left(\frac{D_p}{D_s}\right)}.
    \label{eq3}
    \end{equation}} 

\begin{figure}[h!]
\begin{center}
\includegraphics[angle=0]{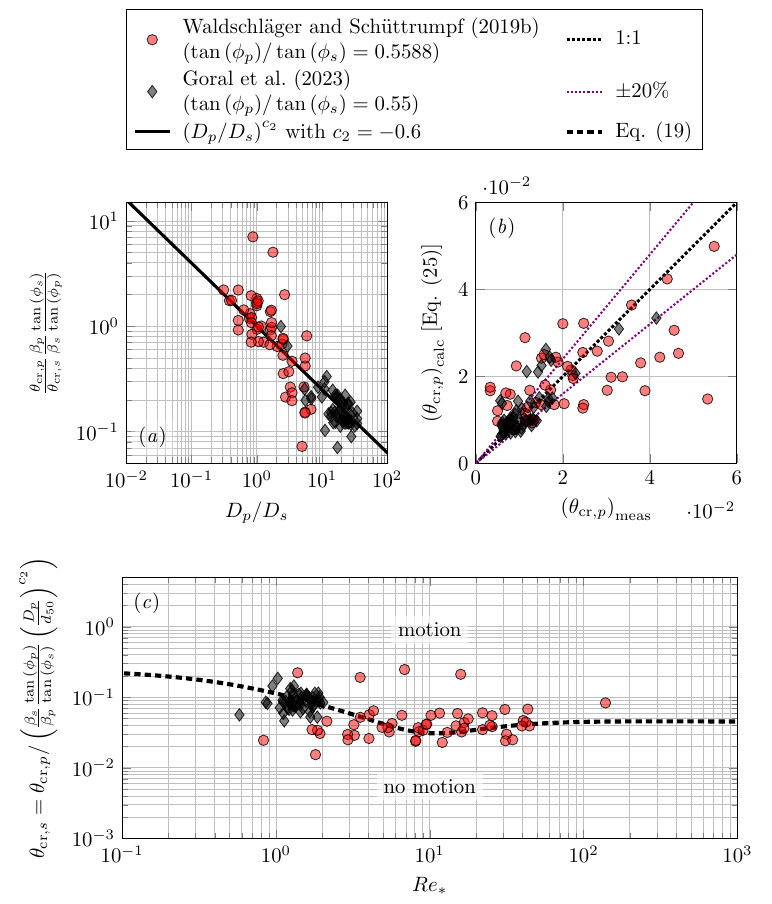}
\end{center}
\caption{Incipient motion of plastic particles: (\textit{a}) hiding exposure-function; (\textit{b})
measured versus calculated plastic Shields parameters
(\textit{c}) classical Shields diagram with modified plastic Shields parameter}
\label{fig4}
\end{figure}

\textcolor{black}{The re-analysis of the two data sets \cite{Waldschlaeger2019,Goral2023} showed that the experimental data scattered around an exponent $c_2= -0.6$. This is exemplified in Fig. \ref{fig4}\textit{a}, where the two sides of Eq. (\ref{eq2}) are plotted against each other, together with the power law relationship.} To reconcile the incipient motion of plastics with classical sediment, the plastic Shields parameter can be normalised as
\begin{equation}
\theta_{\text{cr},s} = \frac{\theta_{\text{cr},p}}{\left( \frac{\beta_s}{\beta_p} \frac{\tan{(\phi_p)}}{\tan{(\phi_s)}}  \left(\frac{D_p}{D_s}  \right)^{c_2} \right)}.
\end{equation}

\renewcommand*{\arraystretch}{1}
\begin{center}
\begin{small}
\begin{table}[h!]
\centering
\caption{Shape factors $\beta_p = (A_\text{proj} D_p)/\mathcal{V}_p$ for selected plastic geometries; note that the diameters are calculated as sphere-volume equivalent diameter $D_p = \sqrt[3]{\frac{6}{\pi} \mathcal{V}_p}$}
\begin{tabular}{c c c c c c c c}
\toprule
\begin{tabular} {@{}c@{}}Shape/ \\ Parameter \end{tabular} & \begin{tabular} {@{}c@{}}Sphere \\ \phantom{P}  \end{tabular} & \begin{tabular} {@{}c@{}}Cylinder \\ \phantom{P} 
\end{tabular} &  \begin{tabular} {@{}c@{}}Cube \\\phantom{P}   \end{tabular} & \begin{tabular} {@{}c@{}}Rectangular prism \\ \phantom{P}  \end{tabular}\\
\midrule
$\mathcal{V}_p$ (m$^3$) &  $\frac{\pi D_p^3}{6}$ & $\frac{\pi D^2 a}{4}$ & $a^3$  & $abc$\\ 
$A_\text{proj}$ (m$^2$) &  $\frac{\pi D_p^2}{4}$ & $a D$ & $a^2$ & $bc$\\
$D_p$ (m) &  $D_p$ & $\sqrt[3]{1.5 D^2 a}$
&$\sqrt[3]{\frac{6}{\pi}}a$ &$\sqrt[3]{\frac{6}{\pi}  abc}$ \\
$\beta_p$ ($-$) & 1.5 & $4\sqrt[3]{1.5a}/(\pi D^{1/3})$ & 1.24 & $\sqrt[3]{\frac{6}{\pi} a b c}/a$\\
\multicolumn{5}{c}{
\hspace{1.7cm}
\begin{tikzpicture}
\draw (0,0) circle (0.75cm);
\draw (-0.75,0) arc (180:360:0.75 and 0.5);
\draw[dashed] (0.75,0) arc (0:180:0.75 and 0.5);
\fill[fill=black] (0,0) circle (1pt);
\draw[dashed] (-0.75,0) -- node[above]{$D_p$} (0.75,0);
\end{tikzpicture}
\hspace{0.1cm}
\begin{tikzpicture}
\draw (-0.75,0) arc (180:360:0.75 and 0.5);
\draw[dashed] (0.75,0) arc (0:180:0.75 and 0.5);
\draw (0.75,1.5) arc (0:360:0.75 and 0.5);
\fill[fill=black] (0,0) circle (1pt);
\draw[dashed] (-0.75,0) -- node[above]{$D$}(0.75,0);
\draw[solid] (0.75,0) -- node[right]{$a$}(0.75,1.5);
\draw[solid] (-.75,0) -- (-0.75,1.5);
\end{tikzpicture}
\hspace{0cm}
\begin{tikzpicture}
\draw (0,1,0) -- node[above]{$a$}(1,1,0) -- (1,0,0) -- (1,0,1) -- (0,0,1) -- (0,1,1) -- cycle;
\draw[line cap=round] (1,1,1) -- (0,1,1) (1,1,1) -- (1,0,1) (1,1,1) -- (1,1,0);
\draw[dashed] (0,0,0) -- (0,1,0);
\draw[dashed] (0,0,0) -- (0,0,1);
\draw[dashed] (0,0,0) -- (1,0,0);
\end{tikzpicture}
\hspace{0.2cm}
\begin{tikzpicture}
\draw (0,1.5,-1) -- node[above]{$a$}(1,1.5,-1) -- node[right]{$b$} (1,0,-1) -- node[right]{$c$}(1,0,1) -- (0,0,1) -- (0,1.5,1) -- cycle;
\draw[line cap=round] (1,1.5,1) -- (0,1.5,1) -- (1,1.5,1) -- (1,0,1) (1,1.5,1) -- (1,1.5,-1);
\draw[dashed] (0,0,-1) -- (0,1.5,-1);
\draw[dashed] (0,0,-1) -- (0,0,1);
\draw[dashed] (0,0,-1) -- (1,0,-1);
\end{tikzpicture}
}\\
\bottomrule
\end{tabular}
\label{table:conditions}
\end{table}
\end{small}
\end{center}

These normalised values are plotted together with the sediment Shields curve [Eq. (\ref{eqSui})] in Fig. \ref{fig4}\textit{c}.
Comparing the Shields parameters of plastics with those of natural sediment (Figs. \ref{fig2} and \ref{fig4}\textit{c}), one can draw the following conclusions on how different physical parameters affect the mobility of plastics: i) the plastic shape can lead to a reduced or increased mobility of plastics compared to sediments ($\beta_s/\beta_p > 1$ or $\beta_s/\beta_p < 1$), ii) the friction term increases the mobility of plastics ($\tan{(\phi_s)}/\tan{(\phi_p)} < 1$), and iii) the plastic size can lead to an increase in mobility if  $D_p > D_s$ (exposure effect), or to a decrease if $D_p < D_s$ (hiding effect). For completeness, a plot of measured versus calculated plastic Shields parameters is shown in Fig. \ref{fig4}\textit{b}. It is seen that there is some data scatter, which is anticipated to be caused by experimental uncertainties and by unaccounted effects deriving from simplifications made to obtain Eq. (\ref{eq13}).
Given that the tested plastic data comprise a variety of different shapes and that the classical Shields diagram for sediments holds a similar level of scatter, the present approach seems reasonable, and it is believed that the proposed framework provides a valuable basis for future research on this topic.

\section{Conclusion}
This work is inspired by the fact that many numerical studies of fluvial plastic transport are treating their particles as ``settled'' or  ``surfaced'' once they reach the bottom boundary or the free-surface in their simulations, revealing an imminent knowledge gap in the understanding of free-surface detachment and bed entrainment of plastic particles. To close this gap, a first principle approach is deployed, enabling the derivation of a novel formulation for free-surface detachment of floating plastics, as well as the creation of a surface detachment/entrapment diagram, which can be regarded as the surface counterpart to the classical Shields diagram. It is acknowledged that some of the parameters relating to surface tension and downpull must be validated or established in physical experiments. A preliminary analysis of the critical detachment velocity, exemplified for cubical and spherical particles, shows that the free-surface acts as sink for microplastics, while larger plastics are likely to detach from the interface. We anticipate that the implementation of this formulation into numerical modelling environments will allow for a more realistic prediction of plastic surface transport processes. \textcolor{black}{Further, there is scope to apply the presented formulations to account for detachment at local disturbances, such as weirs, hydraulic jumps, etc.}

Next, the focus is set on bed entrainment (or incipient motion) of fluvial plastics, and a formulation of the plastic Shields parameter is presented, which is in principle identical to formulations used for natural sediment. Because the establishment of a plastic Shields curve would require dozens of experiments, a time-saving alternative is desired. This is herein achieved by introducing a framework that relates different Shields parameters with one another, thereby allowing to reconcile incipient motion conditions of plastics with those of natural sediments. It is noted that this framework i) is based on first principles, ii) leads to a new expression for the plastic Shields parameter, which represents an expansion of previous approaches used in plastic research, and iii) explicitly accounts for the particle shape. Subsequently, this new framework is tested using two data sets from literature, and a reasonable agreement between predictions and measurements is achieved, while some assumptions need to be scrutinised in the future. Altogether, it is hoped that this work on surface detachment and bed entrainment will be useful for the planning, execution, and analysis of future experiments and simulations of turbulent plastic transport in an environmental fluid mechanics context.

\section*{Open Research Section}
Supporting data used to create figures in this manuscript is associated with the studies from \citeA{Waldschlaeger2019} and \citeA{Goral2023}. These data can be downloaded from
\href{https://pubs.acs.org/doi/abs/10.1021/acs.est.9b05394}{Link 1} and \href{https://doi.org/10.11583/DTU.19762507}{Link 2}. 

\acknowledgments
I would like to thank Deniz Goral and David R. Furhmann for fruitful discussions on bed entrainment of plastics. Further, I acknowledge the exchanges with my PhD students Felipe Condo and Charuni Wickramarachchi, as well as with my colleague  Daniel Valero.

\section*{Notation}
\noindent
The following symbols and abbreviations are used in this manuscript:

\noindent 
Parameters
\vspace{-0.7cm}
\begin{tabbing}
\hspace*{0cm}\=\hspace*{2cm}\=\kill\\
\>$a, b,$ and $c$ \> geometrical dimensions as per Tab. \ref{table:conditions}  (m) \\
\>$A_\text{proj}$ \> projected area for drag and lift (m$^2$) \\
\>$A_{\text{proj},F_w}$ \> projected area for $F_w$  (m$^2$) \\
\>$c_1$ to $c_4$ \>empirical parameters for hiding-exposure function  ($-$) \\
\>$C_D$ \>drag coefficient  ($-$) \\
\>$C_{D,{F_w}}$ \> coefficient for estimating $F_w$ ($-$) \\
\>$C_L$ \>lift coefficient  ($-$) \\
\>$D$ \> diameter of cylinder base (m) \\
\>$D_p$ \> sphere-volume equivalent  diameter (m) \\
\>$R_p$ \> radius of spherical particle (m) \\
\>$d_{50}$ \> median diameter of underlying sediment (m)\\
\>$F_B$ \> buoyancy force (N)\\
\>$F_D$ \> drag force (N)\\
\>$F_L$ \> lift force (N)\\
\>$F_R$ \> resistance force (N)\\
\>$F_W$ \> weight force (N)\\
\>$F_w$ \> vertical water force or downpull force (N)\\
\>$F_\sigma$ \> surface tension force (N)\\
\>$g$ \> gravitational acceleration (m/s$^2$)\\
\>$h_w$ \> submerged depth (m)\\
\>$H$ \> water depth (m)\\
\>$L_\sigma$ \> air-water-plastic interface length (m)\\
\>$Re_p$ \> plastic Reynolds number ($-$)\\
\>$Re_s$ \> sediment Reynolds number ($-$)\\
\>$Re_{*}$ \> shear Reynolds number ($-$)\\
\>$\mathcal{V}$ \> particle volume (m$^3$)\\
\>$\mathcal{V}_{p,w}$ \> submerged volume of plastic particle (m$^3$)\\
\>$u_a$ \> air velocity of the air-superlayer (m/s)\\
\>$u_\text{fs}$ \> fluid surface velocity (m/s)\\
\>$u_p$ \> velocity of surfaced plastic (m/s)\\
\>$u_{\text{cr},p}$ \> critical velocity for surface detachment (m/s)\\
\>$u_{\text{cr}}$ \> critical velocity for bed entrainment (m/s)\\
\>$u_{*}$ \> bed shear velocity (m/s)\\
\>$u_{*,\text{cr}}$ \> critical shear velocity (m/s)\\
\>$w'_\text{rms}$ \> root-mean-square of vertical velocity fluctuations (m/s)\\
\>$x$ \> streamwise coordinate (m)\\
\>$z$ \> vertical coordinate (m)\\
\>$z_0$ \> hydraulic roughness (m)\\
\>$\alpha$ \> $ = u_\text{cr}/u_{*,\text{cr}}$; deviation parameter  ($-$)\\
\>$\beta$ \> $ = (A_\text{proj} D)/\mathcal{V}$; particle shape factor  ($-$)\\
\>$\Gamma$ \> plastic-based E\"otv\"os number ($-$)\\
\>$\theta_{\text{cr}}$ \>  Shields parameter ($-$)\\
\>$\Theta_{\text{cr},p}$ \> dimensionless surface detachment velocity ($-$)\\
\>$\kappa$ \> van Karman constant ($-$)\\
\>$\nu$ \> kinematic water viscosity (m$^2$/s)\\
\>$\rho$ \> density (kg/m$^3$)\\
\>$\sigma$ \> surface tension (N/m)\\
\>$\tau_\text{cr}$ \> critical bed shear stress (N/m$^2$)\\
\>$\phi$ \> friction angle ($^\circ$)\\
\>$\Omega$ \> contact angle ($^\circ$)\\
\end{tabbing}

\noindent 
Indices and abbreviations
\vspace{-0.7cm}
\begin{tabbing}
\hspace*{0cm}\=\hspace*{2cm}\=\hspace*{11.2cm}\=\kill\\
\>$a$ \> air\\
\> cube \> cubical particle\\
\>$\text{cr}$ \> critical\\
\>$\text{max}$ \> maximum\\
\>$p$ \> plastic\\
\>$s$ \> sediment\\
\> sphere \> spherical particle\\
\>$w$ \> water\\
\end{tabbing}

%
%


\appendix
\section{Newton's second law applied to a floating plastic}
\phantomsection

\label{App:vertical}
Here, Newton's second law of motion is applied to a floating plastic in vertical direction. The forces considered are body forces, i.e., gravity, and surface forces,  including pressures and turbulent surface stresses. Newton's second law of motion for a particle is written in $z$-direction as follows
\begin{equation}
\begin{split}
\iiint_{\mathcal{V}_p} \frac{D \left(\rho_p w_p \right) }{Dt} \text{d} \mathcal{V}_p = & \overbrace{\iiint_{\mathcal{V}_p} \rho_p g_z \text{d} \mathcal{V}_p}^{F_W} + \overbrace{\iint_{A_p} - p  n_z  \text{d} A_p}^{F_B}  +  \overbrace{\int_{L_\sigma} \sigma \sin{(\Omega)} \,  \text{d}L_\sigma}^{F_\sigma}  \\  
& + \underbrace{\iint_{A_p} \tau_{zi}  n_i  \, \text{d} A_p   + \iint_{A_p} -p_d  n_z  \text{d} A_p}_{F_w}  
\end{split}
\label{Appforce}
\end{equation}
where $w_p$ is the vertical particle velocity, $g_z = -g$ is the vertical component of gravitational acceleration, $p$ is the hydrostatic pressure, $A_p$ is the surface area of the particle, $\sigma$ is the surface tension, $L_\sigma$ is the interfacial contact length, $\Omega$ the contact angle, and $\sin{(\Omega)}$ accounts for the vertical projection of the surface tension force, $\tau_{zi}$ are the viscous and Reynolds stresses in vertical direction, $n_i$ is the $i$th component of the outward (into the fluid) unit vector perpendicular to the element $\text{d} A_p$ of the particle surface, $p_d$ is the deviatoric pressure, i.e., the deviation from the hydrostatic component, and tensor notation is used with the Einstein convention, which prescribes a summation over each repeated index. Note that it is assumed that the vertical particle velocity is $w_p = 0$, which implies that the left-hand-side of Eq. (\ref{Appforce}) becomes zero. 

Next, the focus is set on the right-hand-side of Eq. (\ref{Appforce}), and the different force terms are simplified as follows.

\begin{itemize}
    \item \textbf{Weight force} $F_W$\\
    The integration of the body force over the volume of the particle yields
\begin{equation}
 F_W =   \iiint_{\mathcal{V}_p} \rho_p g_z \text{d}\mathcal{V}_p = - \rho_p g \mathcal{V}_p,
 \label{AppFW}
\end{equation}
where it was assumed that the particle's density is homogeneous.

   \item \textbf{Buoyancy force} $F_B$\\
The buoyancy term can be simplified as \cite{Crowe2011}
\begin{equation}
F_B =   \iint_{A} - p  n_z  \text{d} A = -  \iiint_{ \mathcal{V}_{p,w}} \frac{\partial p}{\partial z}  \text{d}  \mathcal{V}_{p,w} = -  \mathcal{V}_{p,w} \frac{\partial p}{\partial z}  = \rho_w g \mathcal{V}_{p,w},
 \label{AppFB}
\end{equation}
where the divergence theorem was applied to the submerged volume of the particle $V_{p,w}$. Further, it was assumed that the hydrostatic pressure gradient $\partial p/\partial z = - \rho_w g$ is constant over the submerged volume of the particle. 

   \item \textbf{Surface tension force} $F_\sigma$\\
Surface tension forces appear at the interfacial air-water-plastic contact line of a floating plastic particle \cite{VALERO2022119078}. At this contact line, the interface bends with a certain angle, leading to a vertical component of the surface tension force. Following \citeA{White2016}, the surface tension force is expressed as 
\begin{equation}
F_\sigma = \int_{L_\sigma} \sigma \sin{(\Omega)} \, \text{d}L_\sigma \approx L_\sigma \sigma \sin{(\Omega)}, 
 \label{AppFsigma}
\end{equation}
where we assumed that the contact angle $\Omega$ is representative for the interfacial contact length. 

 \item \textbf{Downpull force} $F_w$\\
The floating particle is exposed to non-hydrostatic pressures and turbulent stresses, exerted by the water phase
\begin{equation}
 F_w = \iint_{A_p} \tau_{zi}  n_i  \, \text{d} A_p   + \iint_{A_p} - p_d  n_z  \text{d} A_p. 
 \label{EqAppFw}
\end{equation}

Eq. (\ref{EqAppFw}) requires precise knowledge on pressures and turbulent stresses acting over the control volume surface, which is not commonly available. To characterise this \mbox{fluid-plastic} interaction, literature approaches distinguish between  quasi-steady drag and unsteady forces, the latter comprising added mass and Basset forces \cite{Crowe2011}. Here, we disregard unsteady forces, and we assume local equilibrium by invoking a steady-state drag assumption
\begin{equation}
 F_w \approx \frac{1}{2} \rho_w C_{D,{F_w}} A_{\text{proj},F_w} \vert w_\text{fs} - w_p  \vert \,  (w_\text{fs} - w_p), 
 \label{EqAppdrag}
\end{equation}
where $C_{D,{F_w}}$ is a (drag-) coefficient, that is able to characterise $F_w$, and $w_\text{fs}$ is the water velocity at the particle's position, i.e., close to the free surface. Solving Eq. (\ref{AppFw}) requires a determination of the coefficient $C_{D,F_w}$, which is dealt with by a sensitivity analysis in $\S$ \ref{DetachmentVel}, acknowledging that  future experimental validation is required.  
To simplify Eq. (\ref{EqAppdrag}), instantaneous water velocities at the free-surface are reformulated using Reynolds decomposition $w_\text{fs} = \overline{w}_\text{fs} + w^\prime_\text{fs}$. Recalling  $w_p = 0$ and assuming that time averaged velocities at the interface are $\overline{w}_\text{fs} \approx 0$, one obtains
\begin{equation}
F_w  = \frac{1}{2} \rho_w C_{D,{F_w}}  A_{\text{proj},F_w} \vert w^{\prime}_\text{fs} \vert w^{\prime}_\text{fs}.
 \label{EqAppdrag1}
\end{equation}

It is important to note that $F_w$ as per Eq. (\ref{EqAppdrag1}) can be directed downwards and upwards, where a downward force would pull the particle back into suspension, while an upward forcing would imply that the particle is pushed upwards, perpendicular to the air-water interface; importantly, in the latter case, the particle would resume its equilibrium position. Hereafter, we are only interested in the case where a particle detaches from the interface, as this may lead to a subsequent change of the particle's position. Similar to \citeA{Ikeda1971} and \citeA{10.1063/1.4955103}, we replace the instantaneous velocity fluctuations $w^\prime$ with their root-mean-square value $w'_\text{rms}$
\begin{equation}
F_w  = - \frac{1}{2} \rho_w C_{D,{F_w}} A_{\text{proj},F_w} w'^2_\text{rms},
 \label{AppFw}
\end{equation}
where a negative sign was included to indicate downward forcing. 
\end{itemize}

Combining Eqns. (\ref{AppFW}), (\ref{AppFB}), (\ref{AppFsigma}), and (\ref{AppFw}), and assuming no vertical acceleration, it is  obtained
\begin{equation}
\underbrace{\vphantom{\frac{1}{2}}\rho_w g \mathcal{V}_{p,w}}_{F_B} + \underbrace{\vphantom{\frac{1}{2}}L_\sigma \, \sigma \, \sin{(\Omega)}}_{F_\sigma} =  \underbrace{\vphantom{\frac{1}{2}}\rho_p g\mathcal{V}_p}_{F_W} + \underbrace{\frac{1}{2} \rho_w C_{D,{F_w}} A_{\text{proj},F_w} \, w'^2_\text{rms}}_{F_w}.
\end{equation}

\color{black}
\section{Detachment of spherical particles}
\begin{figure}[h!]
\includegraphics[angle=0]{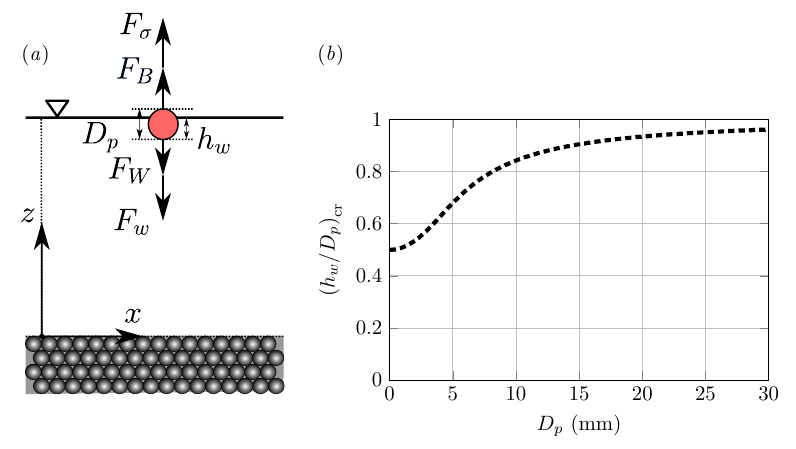}
\caption{Detachment of spherical plastic particles: (\textit{a}) conceptual sketch, including forces acting on the floating sphere; (\textit{b}) relative submergence as function of the particle diameter $D_p$}
\label{figA1}
\end{figure}

\phantomsection
\label{AppA}
To formulate the detachment condition for spherical particles, let us consider a floating plastic sphere shown in Fig. \ref{figA1}\textit{a}, subject to buoyancy $F_B$, surface tension $F_\sigma$, weight force $F_W$, and downpull $F_w$. Here, $D_p$ and $h_w$ are the diameter and the submerged depth, respectively. Considering the forces acting on the spherical particle, it becomes clear that the interfacial contact length $L_\sigma$ is a function of $h_w$, and that i) $L_\sigma \rightarrow 0$ as $h_w \rightarrow D_p$, and ii)  $L_\sigma \rightarrow \text{max}$ as $h_w \rightarrow 0.5 D_p$. More generally, the relative and absolute contributions of $F_B$, $F_\sigma$, and $F_W$ depend on the submergence of the spherical particle, similarly discussed for plastic cups in \citeA[Appendix B]{VALERO2022119078}, while the projected area $A_{\text{proj},F_w} = \pi R_p^2$ in Eq. (\ref{eqparticlevelocity1}) is perpendicular to $F_w$, which suggests, in a first approximation, that $F_w$ does not depend on $h_w$. As such, the detachment condition for spherical particles is postulated as

\begin{equation}
h_{\text{cr},w} = \arg \max_{h_w} \left( \rho_w g \mathcal{V}_{p,w} -  \rho_p g\mathcal{V}_p + L_\sigma \, \sigma \, \sin{(\Omega)}       \right),
\label{hcrsphere}
\end{equation}
which implies that $F_w$ has to overcome the maximum vertical force, formed by $F_B$, $F_\sigma$, and $F_W$, for the particle to be detached. In Eq. (\ref{hcrsphere}), the submerged volume of the spherical particle is $\mathcal{V}_{p,w} = \frac{\pi h_{\text{cr},w}^2}{3}(3 R_p - h_{\text{cr},w})$, the total volume  $\mathcal{V}_p = \frac{4}{3} \pi R_p^3$, and the interfacial contact length $L_\sigma =2 \pi \sqrt{2 \, h_{w} \, R_p - h_{w}^2}$, with $R_p = D_p/2$ being the radius. Figure \ref{figA1}\textit{b} shows the detachment condition $(h_w/D_p)_\text{cr}$ for spheres with diameters ranging from $D_p = 0$ to 30 mm, which was evaluated using Eq. (\ref{hcrsphere}), further assuming $\sigma = 0.072$ N/m and $\Omega = 105^\circ$. Because weight and buoyant forces are negligible for small particles, $(h_w/D_p)_\text{cr} \approx 0.5$ for $D_p < 1$ mm, and $(h_w/D_p)_\text{cr}$ increases towards unity with increasing $D_p$. Notably, the results shown in Fig. \ref{figA1}\textit{b} are independent of the particle density $\rho_p$.

\begin{figure}[h!]
\begin{center}
\includegraphics[angle=0]{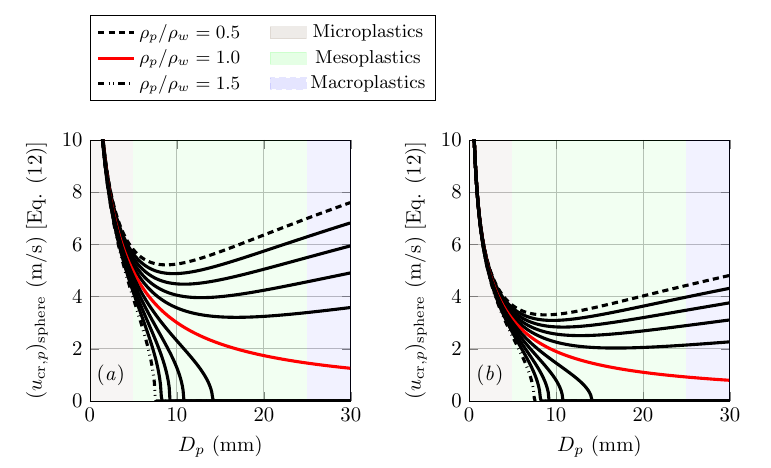}
\end{center}
\caption{Critical particle detachment velocities for spherical particles:  (\textit{a}) dimensional detachment velocity versus particle diameter for a coefficient $C_{D,{F_w}} = 2$; the simulated channel is characterised by $H/z_0 = 100$; (\textit{b}) same as (\textit{a}) but with $C_{D,{F_w}} = 5$}
\label{figA2}
\end{figure}

Having determined $h_{\text{cr},w}$ as per Fig. \ref{figA1}\textit{b}, detachment velocities $u_{\text{cr},p}$ for spherical particles can be evaluated using Eq. (\ref{eqverticalforce0}). Here, previously used parameter values were adopted, including $H/z_0 = 100$, $C_{D,{F_w}} = 2$ and $5$, $\sigma = 0.072$ N/m,  $\Omega = 105^\circ$, whereas $\left(\mathcal{V}_{p,w} \right)_\text{cr}$, $L_{\text{cr},\sigma}$, and $\left( \frac{A_{\text{proj},a}}{A_{\text{proj},w}} \right)_\text{cr}$ are related to $h_{\text{cr},w}$ through 
\begin{equation}
\left(\mathcal{V}_{p,w} \right)_\text{cr} =   \frac{\pi \, h_{\text{cr},w}^2}{3} (3 R_p - h_{\text{cr},w}),
 \label{eqA2}
\end{equation}
\begin{equation}
 L_{\text{cr},\sigma} = 2 \pi \sqrt{2 \, h_{\text{cr},w} \, R_p - h_{\text{cr},w}^2},
 \label{eqA3}
\end{equation}
\begin{equation}
\left( \frac{A_{\text{proj},a}}{A_{\text{proj},w}} \right)_\text{cr} = \frac{\pi R_p^2}{R_p^2 \arccos \left(1 - \frac{h_{\text{cr},w}}{R_p}  \right) - (R_p-h_{\text{cr},w}) \sqrt{R_p^2 - \left(R_p - h_{\text{cr},w} \right)^2}}-1.
 \label{eqA4}
\end{equation}

The results of this analysis are shown in Fig. \ref{figA2} for densities ranging from $0.5 \leq \rho_p/\rho_w \leq 1.5$, separated by increments of 0.1. Overall, the trends are very similar to the detachment velocities of cubical particles, compare Fig. \ref{fig2c}, with physical implications as discussed,  while critical velocities of spherical particles are shifted to slightly smaller values when compared to cubical particles, which is due to shorter interfacial contact lengths $L_{\text{cr},\sigma}$. Next, dimensionless detachment velocities $(\Theta_\text{cr,p})_\text{sphere}$ are calculated using Eq. (\ref{dimensionless1}) in combination with Eqns. (\ref{hcrsphere}), (\ref{eqA2}), (\ref{eqA3}), and (\ref{eqA4}), and have been shown in Fig. \ref{fig2c}\textit{c}. At last, it is noteworthy mentioning that  detachment conditions for other closed-shaped particles, such  as cylinders,  rectangular prisms, and ellipsoids, can easily be deduced using the detachment condition formulated in Eq. (\ref{hcrsphere}).

\section{Plastic Shields parameter}
\phantomsection
\label{ShieldsSediment}
In this section, a derivation of the plastic Shields parameter is presented, which closely follows previous works on the incipient motion of natural sediment by \citeA{Wiberg1987}. Our starting point is the streamwise force balance on the deposited plastic at the onset of motion, i.e., Eq. (\ref{eq1}), which is repeated for convenience
\begin{equation}
 \frac{1}{2} \rho_w \, u_\text{cr}^2  C_{D,p}  A_\text{proj} =  \left( (\rho_p -\rho_w) g  \mathcal{V}_{p} - \frac{1}{2} \rho_w u_\text{cr}^2 C_{L,p} A_\text{proj} \right) \tan{(\phi_p)}.
 \label{app1}
\end{equation}

Rearranging Eq. (\ref{app1})
\begin{equation}
\frac{1}{2} \rho_w \, u_\text{cr}^2   A_\text{proj} (C_{D,p} + C_{L,p} \tan{(\phi_p)}) =  (\rho_p -\rho_w) g  \mathcal{V}_{p}
\tan{(\phi_p)},
\end{equation}
and dividing by $\frac{1}{2} A_\text{proj} (C_{D,p} + C_{L,p} \tan{(\phi_p)}) (\rho_p - \rho_w) g D_p$ yields 
\begin{equation}
\frac{u_\text{cr}^2  }{ \left(\frac{\rho_p - \rho_w}{\rho_w} \right) g D_p} = \frac{2 \mathcal{V}_{p}}{A_\text{proj} D_p} \frac{
\tan{(\phi_p)}}{  C_{D,p} + C_{L,p} \tan{(\phi_p)}}
 \label{app3}
\end{equation}

Next, a shape factor $\beta_p = (A_\text{proj} D_p)/\mathcal{V}_p$ and a parameter $\alpha = u_\text{cr}/u_{*,\text{cr}}$ are introduced, where the latter accounts for deviations between the critical velocity and the shear velocity. The critical shear velocity is defined as $u_{*,\text{cr}} = \sqrt{\tau_{\text{cr},p}/ \rho_w}$, with $\tau_{\text{cr},p}$ being the critical bed shear stress. Substituting $\alpha_p$ and $\beta_p$ into Eq.  (\ref{app3}) leads to the following expression of the plastic Shields parameter
\begin{equation}
\theta_{\text{cr},p} = \frac{u_{*,\text{cr}}^2}{ \left(\frac{\rho_p - \rho_w}{\rho_w} \right) g D_p} = \frac{2}{\beta_p} \, \frac{1}{\alpha_p^2} \,
\frac{
\tan{(\phi_p)}}{  C_{D,p} + C_{L,p} \tan{(\phi_p)}},
\label{eqShieldsustar}
\end{equation}
which can be slightly reformulated using the definition of the shear velocity 
\begin{equation}
\theta_{\text{cr},p} = \frac{\tau_{\text{cr},p}}{ \left(\rho_p - \rho_w \right) g D_p} = \frac{2}{\beta_p} \, \frac{1}{\alpha_p^2} \,
\frac{
\tan{(\phi_p)}}{  C_{D,p} + C_{L,p} \tan{(\phi_p)}}.
\end{equation}

\bibliography{bibliography}

\begin{thebibliography}{}

\bibitem [\protect \citeauthoryear {%
Ali%
\ \BBA {} Dey%
}{%
Ali%
\ \BBA {} Dey%
}{%
{\protect \APACyear {2016}}%
}]{%
10.1063/1.4955103}
\APACinsertmetastar {%
10.1063/1.4955103}%
\begin{APACrefauthors}%
Ali, S\BPBI Z.%
\BCBT {}\ \BBA {} Dey, S.%
\end{APACrefauthors}%
\unskip\
\newblock
\APACrefYearMonthDay{2016}{07}{}.
\newblock
{\BBOQ}\APACrefatitle {Hydrodynamics of sediment threshold} {Hydrodynamics of sediment threshold}.{\BBCQ}
\newblock
\APACjournalVolNumPages{Physics of Fluids}{28}{7}{075103}.
\PrintBackRefs{\CurrentBib}

\bibitem [\protect \citeauthoryear {%
Bagheri%
\ \BBA {} Bonadonna%
}{%
Bagheri%
\ \BBA {} Bonadonna%
}{%
{\protect \APACyear {2016}}%
}]{%
bagh_2016}
\APACinsertmetastar {%
bagh_2016}%
\begin{APACrefauthors}%
Bagheri, G.%
\BCBT {}\ \BBA {} Bonadonna, C.%
\end{APACrefauthors}%
\unskip\
\newblock
\APACrefYearMonthDay{2016}{}{}.
\newblock
{\BBOQ}\APACrefatitle {{On the drag of freely falling non-spherical particles}} {{On the drag of freely falling non-spherical particles}}.{\BBCQ}
\newblock
\APACjournalVolNumPages{Powder Technology}{301}{}{526--544}.
\PrintBackRefs{\CurrentBib}

\bibitem [\protect \citeauthoryear {%
Chubarenko%
, Bagaev%
, Zobkov%
\BCBL {}\ \BBA {} Esiukova%
}{%
Chubarenko%
\ \protect \BOthers {.}}{%
{\protect \APACyear {2016}}%
}]{%
CHUBARENKO2016105}
\APACinsertmetastar {%
CHUBARENKO2016105}%
\begin{APACrefauthors}%
Chubarenko, I.%
, Bagaev, A.%
, Zobkov, M.%
\BCBL {}\ \BBA {} Esiukova, E.%
\end{APACrefauthors}%
\unskip\
\newblock
\APACrefYearMonthDay{2016}{}{}.
\newblock
{\BBOQ}\APACrefatitle {On some physical and dynamical properties of microplastic particles in marine environment} {On some physical and dynamical properties of microplastic particles in marine environment}.{\BBCQ}
\newblock
\APACjournalVolNumPages{Marine Pollution Bulletin}{108}{1}{105-112}.
\PrintBackRefs{\CurrentBib}

\bibitem [\protect \citeauthoryear {%
Cowger%
, Gray%
, Guilinger%
, Fong%
\BCBL {}\ \BBA {} Waldschl\"ager%
}{%
Cowger%
\ \protect \BOthers {.}}{%
{\protect \APACyear {2021}}%
}]{%
cowger2021concentration}
\APACinsertmetastar {%
cowger2021concentration}%
\begin{APACrefauthors}%
Cowger, W.%
, Gray, A\BPBI B.%
, Guilinger, J\BPBI J.%
, Fong, B.%
\BCBL {}\ \BBA {} Waldschl\"ager, K.%
\end{APACrefauthors}%
\unskip\
\newblock
\APACrefYearMonthDay{2021}{}{}.
\newblock
{\BBOQ}\APACrefatitle {Concentration depth profiles of microplastic particles in river flow and implications for surface sampling} {Concentration depth profiles of microplastic particles in river flow and implications for surface sampling}.{\BBCQ}
\newblock
\APACjournalVolNumPages{Environmental Science \& Technology}{55}{9}{6032--6041}.
\PrintBackRefs{\CurrentBib}

\bibitem [\protect \citeauthoryear {%
Crowe%
, Schwarzkopf%
, Sommerfeld%
\BCBL {}\ \BBA {} Tsuji%
}{%
Crowe%
\ \protect \BOthers {.}}{%
{\protect \APACyear {2011}}%
}]{%
Crowe2011}
\APACinsertmetastar {%
Crowe2011}%
\begin{APACrefauthors}%
Crowe, C\BPBI T.%
, Schwarzkopf, J\BPBI D.%
, Sommerfeld, M.%
\BCBL {}\ \BBA {} Tsuji, Y.%
\end{APACrefauthors}%
\unskip\
\newblock
\APACrefYear{2011}.
\newblock
\APACrefbtitle {Multiphase flows with droplets and particles} {Multiphase flows with droplets and particles}\ [Book].
\newblock
\APACaddressPublisher{}{CRC press}.
\PrintBackRefs{\CurrentBib}

\bibitem [\protect \citeauthoryear {%
de Luna%
\ \protect \BOthers {.}}{%
de Luna%
\ \protect \BOthers {.}}{%
{\protect \APACyear {2014}}%
}]{%
deLuna2014}
\APACinsertmetastar {%
deLuna2014}%
\begin{APACrefauthors}%
de Luna, M\BPBI S.%
, Galizia, M.%
, Wojnarowicz, J.%
, Rosa, R.%
, Lojkowski, W.%
, Leonelli, C.%
\BDBL {}Filippone, G.%
\end{APACrefauthors}%
\unskip\
\newblock
\APACrefYearMonthDay{2014}{}{}.
\newblock
{\BBOQ}\APACrefatitle {Dispersing hydrophilic nanoparticles in hydrophobic polymers: HDPE/ZnO nanocomposites by a novel template-based approach} {Dispersing hydrophilic nanoparticles in hydrophobic polymers: Hdpe/zno nanocomposites by a novel template-based approach}.{\BBCQ}
\newblock
\APACjournalVolNumPages{eXPRESS Polymer Letters}{8}{5}{362--372}.
\PrintBackRefs{\CurrentBib}

\bibitem [\protect \citeauthoryear {%
Dey%
}{%
Dey%
}{%
{\protect \APACyear {2014}}%
}]{%
dey2014fluvial}
\APACinsertmetastar {%
dey2014fluvial}%
\begin{APACrefauthors}%
Dey, S.%
\end{APACrefauthors}%
\unskip\
\newblock
\APACrefYear{2014}.
\newblock
\APACrefbtitle {Fluvial hydrodynamics} {Fluvial hydrodynamics}.
\newblock
\APACaddressPublisher{}{Springer}.
\PrintBackRefs{\CurrentBib}

\bibitem [\protect \citeauthoryear {%
Dey%
, Ali%
\BCBL {}\ \BBA {} Padhi%
}{%
Dey%
\ \protect \BOthers {.}}{%
{\protect \APACyear {2020}}%
}]{%
DeyLift}
\APACinsertmetastar {%
DeyLift}%
\begin{APACrefauthors}%
Dey, S.%
, Ali, S\BPBI Z.%
\BCBL {}\ \BBA {} Padhi, E.%
\end{APACrefauthors}%
\unskip\
\newblock
\APACrefYearMonthDay{2020}{}{}.
\newblock
{\BBOQ}\APACrefatitle {Hydrodynamic lift on sediment particles at entrainment: present status and its prospect} {Hydrodynamic lift on sediment particles at entrainment: present status and its prospect}.{\BBCQ}
\newblock
\APACjournalVolNumPages{J. Hydraul. Engng.}{146}{6}{}.
\PrintBackRefs{\CurrentBib}

\bibitem [\protect \citeauthoryear {%
Dey%
, Zeeshan~Ali%
\BCBL {}\ \BBA {} Padhi%
}{%
Dey%
\ \protect \BOthers {.}}{%
{\protect \APACyear {2019}}%
}]{%
Dey2019}
\APACinsertmetastar {%
Dey2019}%
\begin{APACrefauthors}%
Dey, S.%
, Zeeshan~Ali, S.%
\BCBL {}\ \BBA {} Padhi, E.%
\end{APACrefauthors}%
\unskip\
\newblock
\APACrefYearMonthDay{2019}{}{}.
\newblock
{\BBOQ}\APACrefatitle {{Terminal fall velocity: The legacy of Stokes from the perspective of fluvial hydraulics}} {{Terminal fall velocity: The legacy of Stokes from the perspective of fluvial hydraulics}}{\BBCQ}\ [Journal Article].
\newblock
\APACjournalVolNumPages{Proc Math Phys Eng Sci}{475}{2228}{20190277}.
\PrintBackRefs{\CurrentBib}

\bibitem [\protect \citeauthoryear {%
Dioguardi%
, Mele%
\BCBL {}\ \BBA {} Dellino%
}{%
Dioguardi%
\ \protect \BOthers {.}}{%
{\protect \APACyear {2018}}%
}]{%
Dio_2018}
\APACinsertmetastar {%
Dio_2018}%
\begin{APACrefauthors}%
Dioguardi, F.%
, Mele, D.%
\BCBL {}\ \BBA {} Dellino, P.%
\end{APACrefauthors}%
\unskip\
\newblock
\APACrefYearMonthDay{2018}{}{}.
\newblock
{\BBOQ}\APACrefatitle {{A new one-equation model of fluid drag for irregularly shaped particles valid over a wide range of Reynolds number}} {{A new one-equation model of fluid drag for irregularly shaped particles valid over a wide range of Reynolds number}}.{\BBCQ}
\newblock
\APACjournalVolNumPages{Journal of Geophysical Research: Solid Earth}{123}{1}{144--156}.
\PrintBackRefs{\CurrentBib}

\bibitem [\protect \citeauthoryear {%
{Diversified Enterprises}%
}{%
{Diversified Enterprises}%
}{%
{\protect \APACyear {2009}}%
{\protect \APACexlab {{\protect \BCnt {1}}}}}]{%
ACCU2009PE}
\APACinsertmetastar {%
ACCU2009PE}%
\begin{APACrefauthors}%
{Diversified Enterprises}.%
\end{APACrefauthors}%
\unskip\
\newblock
\APACrefYearMonthDay{2009{\protect \BCnt {1}}}{}{}.
\newblock
\APACrefbtitle {Surface Energy Data for PE: Polyethylene, CAS \# 9002-88-4.} {Surface energy data for pe: Polyethylene, cas \# 9002-88-4.}
\newblock
\begin{APACrefURL} \url{https://www.accudynetest.com/polymer_surface_data/polyethylene.pdf} \end{APACrefURL}
\PrintBackRefs{\CurrentBib}

\bibitem [\protect \citeauthoryear {%
{Diversified Enterprises}%
}{%
{Diversified Enterprises}%
}{%
{\protect \APACyear {2009}}%
{\protect \APACexlab {{\protect \BCnt {2}}}}}]{%
ACCU2009PP}
\APACinsertmetastar {%
ACCU2009PP}%
\begin{APACrefauthors}%
{Diversified Enterprises}.%
\end{APACrefauthors}%
\unskip\
\newblock
\APACrefYearMonthDay{2009{\protect \BCnt {2}}}{}{}.
\newblock
\APACrefbtitle {Surface Energy Data for PP: Polypropylene, CAS \#s 9003-08- 0 (atactic) and 25085-53-4 (isotactic).} {Surface energy data for pp: Polypropylene, cas \#s 9003-08- 0 (atactic) and 25085-53-4 (isotactic).}
\newblock
\begin{APACrefURL} \url{https://www.accudynetest.com/polymer_surface_data/polypropylene.pdf} \end{APACrefURL}
\PrintBackRefs{\CurrentBib}

\bibitem [\protect \citeauthoryear {%
Eagleson%
\ \BBA {} Dean%
}{%
Eagleson%
\ \BBA {} Dean%
}{%
{\protect \APACyear {1961}}%
}]{%
doi:10.1061/TACEAT.0008054}
\APACinsertmetastar {%
doi:10.1061/TACEAT.0008054}%
\begin{APACrefauthors}%
Eagleson, P\BPBI S.%
\BCBT {}\ \BBA {} Dean, R\BPBI G.%
\end{APACrefauthors}%
\unskip\
\newblock
\APACrefYearMonthDay{1961}{}{}.
\newblock
{\BBOQ}\APACrefatitle {Wave-Induced Motion of Bottom Sediment Particles} {Wave-induced motion of bottom sediment particles}.{\BBCQ}
\newblock
\APACjournalVolNumPages{Transactions of the American Society of Civil Engineers}{126}{1}{1162-1185}.
\PrintBackRefs{\CurrentBib}

\bibitem [\protect \citeauthoryear {%
Feehan%
, McCoy%
, Scheingross%
\BCBL {}\ \BBA {} Gardner%
}{%
Feehan%
\ \protect \BOthers {.}}{%
{\protect \APACyear {2023}}%
}]{%
https://doi.org/10.1029/2023JF007162}
\APACinsertmetastar {%
https://doi.org/10.1029/2023JF007162}%
\begin{APACrefauthors}%
Feehan, S\BPBI A.%
, McCoy, S\BPBI W.%
, Scheingross, J\BPBI S.%
\BCBL {}\ \BBA {} Gardner, M\BPBI H.%
\end{APACrefauthors}%
\unskip\
\newblock
\APACrefYearMonthDay{2023}{}{}.
\newblock
{\BBOQ}\APACrefatitle {Quantifying Variability of Incipient-Motion Thresholds in Gravel-Bedded Rivers Using a Grain-Scale Force-Balance Model} {Quantifying variability of incipient-motion thresholds in gravel-bedded rivers using a grain-scale force-balance model}.{\BBCQ}
\newblock
\APACjournalVolNumPages{Journal of Geophysical Research: Earth Surface}{128}{9}{e2023JF007162}.
\PrintBackRefs{\CurrentBib}

\bibitem [\protect \citeauthoryear {%
Goral%
\ \protect \BOthers {.}}{%
Goral%
\ \protect \BOthers {.}}{%
{\protect \APACyear {2023}}%
}]{%
Goral2023}
\APACinsertmetastar {%
Goral2023}%
\begin{APACrefauthors}%
Goral, K\BPBI D.%
, Guler, H\BPBI G.%
, Larsen, B\BPBI E.%
, Carstensen, S.%
, Christensen, E\BPBI D.%
, Kerpen, N\BPBI B\BPBI B.%
\BDBL {}Fuhrman, D\BPBI R\BPBI R.%
\end{APACrefauthors}%
\unskip\
\newblock
\APACrefYearMonthDay{2023}{2023 JUN 16}{}.
\newblock
{\BBOQ}\APACrefatitle {Shields diagram and the incipient motion of microplastic particles} {Shields diagram and the incipient motion of microplastic particles}.{\BBCQ}
\newblock
\APACjournalVolNumPages{Environ. Sci. Technol.}{}{}{}.
\PrintBackRefs{\CurrentBib}

\bibitem [\protect \citeauthoryear {%
Ikeda%
}{%
Ikeda%
}{%
{\protect \APACyear {1971}}%
}]{%
Ikeda1971}
\APACinsertmetastar {%
Ikeda1971}%
\begin{APACrefauthors}%
Ikeda, S.%
\end{APACrefauthors}%
\unskip\
\newblock
\APACrefYearMonthDay{1971}{}{}.
\newblock
{\BBOQ}\APACrefatitle {Some studies on the mechanics of bed load transport, Proceedings of the Japan Society of Civil Engineers} {Some studies on the mechanics of bed load transport, proceedings of the japan society of civil engineers}.{\BBCQ}.
\PrintBackRefs{\CurrentBib}

\bibitem [\protect \citeauthoryear {%
Kirchner%
, Dietrich%
, Iseya%
\BCBL {}\ \BBA {} Ikeda%
}{%
Kirchner%
\ \protect \BOthers {.}}{%
{\protect \APACyear {1990}}%
}]{%
https://doi.org/10.1111/j.1365-3091.1990.tb00627.x}
\APACinsertmetastar {%
https://doi.org/10.1111/j.1365-3091.1990.tb00627.x}%
\begin{APACrefauthors}%
Kirchner, J\BPBI W.%
, Dietrich, W\BPBI E.%
, Iseya, F.%
\BCBL {}\ \BBA {} Ikeda, H.%
\end{APACrefauthors}%
\unskip\
\newblock
\APACrefYearMonthDay{1990}{}{}.
\newblock
{\BBOQ}\APACrefatitle {The variability of critical shear stress, friction angle, and grain protrusion in water-worked sediments} {The variability of critical shear stress, friction angle, and grain protrusion in water-worked sediments}.{\BBCQ}
\newblock
\APACjournalVolNumPages{Sedimentology}{37}{4}{647-672}.
\PrintBackRefs{\CurrentBib}

\bibitem [\protect \citeauthoryear {%
Lofty%
, Valero%
, Wilson%
, Franca%
\BCBL {}\ \BBA {} Ouro%
}{%
Lofty%
\ \protect \BOthers {.}}{%
{\protect \APACyear {2023}}%
}]{%
LOFTY2023120329}
\APACinsertmetastar {%
LOFTY2023120329}%
\begin{APACrefauthors}%
Lofty, J.%
, Valero, D.%
, Wilson, C.%
, Franca, M.%
\BCBL {}\ \BBA {} Ouro, P.%
\end{APACrefauthors}%
\unskip\
\newblock
\APACrefYearMonthDay{2023}{}{}.
\newblock
{\BBOQ}\APACrefatitle {{Microplastic and natural sediment in bed load saltation: Material does not dictate the fate}} {{Microplastic and natural sediment in bed load saltation: Material does not dictate the fate}}.{\BBCQ}
\newblock
\APACjournalVolNumPages{Water Research}{243}{}{120329}.
\PrintBackRefs{\CurrentBib}

\bibitem [\protect \citeauthoryear {%
Miller%
\ \BBA {} Byrne%
}{%
Miller%
\ \BBA {} Byrne%
}{%
{\protect \APACyear {1966}}%
}]{%
https://doi.org/10.1111/j.1365-3091.1966.tb01897.x}
\APACinsertmetastar {%
https://doi.org/10.1111/j.1365-3091.1966.tb01897.x}%
\begin{APACrefauthors}%
Miller, R\BPBI L.%
\BCBT {}\ \BBA {} Byrne, R\BPBI J.%
\end{APACrefauthors}%
\unskip\
\newblock
\APACrefYearMonthDay{1966}{}{}.
\newblock
{\BBOQ}\APACrefatitle {The angle of response for a single grain on a fixed rough bed} {The angle of response for a single grain on a fixed rough bed}.{\BBCQ}
\newblock
\APACjournalVolNumPages{Sedimentology}{6}{4}{303-314}.
\PrintBackRefs{\CurrentBib}

\bibitem [\protect \citeauthoryear {%
Nezu%
\ \BBA {} Nakagawa%
}{%
Nezu%
\ \BBA {} Nakagawa%
}{%
{\protect \APACyear {1993}}%
}]{%
nezu1993turbulence}
\APACinsertmetastar {%
nezu1993turbulence}%
\begin{APACrefauthors}%
Nezu, I.%
\BCBT {}\ \BBA {} Nakagawa, H.%
\end{APACrefauthors}%
\unskip\
\newblock
\APACrefYearMonthDay{1993}{}{}.
\newblock
{\BBOQ}\APACrefatitle {Turbulence in open-channel flows} {Turbulence in open-channel flows}.{\BBCQ}
\newblock
\BIn{} \APACrefbtitle {Ser., A} {Ser., a}\ (\BPGS\ 1--281).
\PrintBackRefs{\CurrentBib}

\bibitem [\protect \citeauthoryear {%
Rohais%
\ \protect \BOthers {.}}{%
Rohais%
\ \protect \BOthers {.}}{%
{\protect \APACyear {2024}}%
}]{%
ROHAIS2024104822}
\APACinsertmetastar {%
ROHAIS2024104822}%
\begin{APACrefauthors}%
Rohais, S.%
, Armitage, J\BPBI J.%
, Romero-Sarmiento, M\BHBI F.%
, Pierson, J\BHBI L.%
, Teles, V.%
, Bauer, D.%
\BDBL {}Pelerin, M.%
\end{APACrefauthors}%
\unskip\
\newblock
\APACrefYearMonthDay{2024}{}{}.
\newblock
{\BBOQ}\APACrefatitle {A source-to-sink perspective of an anthropogenic marker: A first assessment of microplastics concentration, pathways, and accumulation across the environment} {A source-to-sink perspective of an anthropogenic marker: A first assessment of microplastics concentration, pathways, and accumulation across the environment}.{\BBCQ}
\newblock
\APACjournalVolNumPages{Earth-Science Reviews}{254}{}{104822}.
\PrintBackRefs{\CurrentBib}

\bibitem [\protect \citeauthoryear {%
Sui%
, Staunstrup%
, Carstensen%
\BCBL {}\ \BBA {} Fuhrman%
}{%
Sui%
\ \protect \BOthers {.}}{%
{\protect \APACyear {2021}}%
}]{%
Sui2021}
\APACinsertmetastar {%
Sui2021}%
\begin{APACrefauthors}%
Sui, T.%
, Staunstrup, L\BPBI H.%
, Carstensen, S.%
\BCBL {}\ \BBA {} Fuhrman, D\BPBI R.%
\end{APACrefauthors}%
\unskip\
\newblock
\APACrefYearMonthDay{2021}{}{}.
\newblock
{\BBOQ}\APACrefatitle {Span shoulder migration in three-dimensional current-induced scour beneath submerged pipelines} {Span shoulder migration in three-dimensional current-induced scour beneath submerged pipelines}.{\BBCQ}
\newblock
\APACjournalVolNumPages{Coast. Engng.}{164}{}{}.
\PrintBackRefs{\CurrentBib}

\bibitem [\protect \citeauthoryear {%
Valero%
, Belay%
, Moreno-Rodenas%
, Kramer%
\BCBL {}\ \BBA {} Franca%
}{%
Valero%
\ \protect \BOthers {.}}{%
{\protect \APACyear {2022}}%
}]{%
VALERO2022119078}
\APACinsertmetastar {%
VALERO2022119078}%
\begin{APACrefauthors}%
Valero, D.%
, Belay, B\BPBI S.%
, Moreno-Rodenas, A.%
, Kramer, M.%
\BCBL {}\ \BBA {} Franca, M\BPBI J.%
\end{APACrefauthors}%
\unskip\
\newblock
\APACrefYearMonthDay{2022}{}{}.
\newblock
{\BBOQ}\APACrefatitle {The key role of surface tension in the transport and quantification of plastic pollution in rivers} {The key role of surface tension in the transport and quantification of plastic pollution in rivers}.{\BBCQ}
\newblock
\APACjournalVolNumPages{Water Research}{226}{}{119078}.
\PrintBackRefs{\CurrentBib}

\bibitem [\protect \citeauthoryear {%
Van~Melkebeke%
, Janssen%
\BCBL {}\ \BBA {} De~Meester%
}{%
Van~Melkebeke%
\ \protect \BOthers {.}}{%
{\protect \APACyear {2020}}%
}]{%
Melk_2020}
\APACinsertmetastar {%
Melk_2020}%
\begin{APACrefauthors}%
Van~Melkebeke, M.%
, Janssen, C.%
\BCBL {}\ \BBA {} De~Meester, S.%
\end{APACrefauthors}%
\unskip\
\newblock
\APACrefYearMonthDay{2020}{}{}.
\newblock
{\BBOQ}\APACrefatitle {{Characteristics and sinking behavior of typical microplastics including the potential effect of biofouling: Implications for remediation}} {{Characteristics and sinking behavior of typical microplastics including the potential effect of biofouling: Implications for remediation}}{\BBCQ}\ [Journal Article].
\newblock
\APACjournalVolNumPages{Environ Sci Technol}{54}{14}{8668-8680}.
\PrintBackRefs{\CurrentBib}

\bibitem [\protect \citeauthoryear {%
Vlaeva%
\ \protect \BOthers {.}}{%
Vlaeva%
\ \protect \BOthers {.}}{%
{\protect \APACyear {2012}}%
}]{%
Vlaeva2012}
\APACinsertmetastar {%
Vlaeva2012}%
\begin{APACrefauthors}%
Vlaeva, I.%
, Yovcheva, T.%
, Viraneva, A.%
, Kitova, S.%
, Exner, G.%
, Guzhova, A.%
\BCBL {}\ \BBA {} Galikhanov, M.%
\end{APACrefauthors}%
\unskip\
\newblock
\APACrefYearMonthDay{2012}{}{}.
\newblock
{\BBOQ}\APACrefatitle {Contact angle analysis of corona treated polypropylene films} {Contact angle analysis of corona treated polypropylene films}.{\BBCQ}
\newblock
\APACjournalVolNumPages{Journal of Physics: Conference Series}{398}{012054}{}.
\PrintBackRefs{\CurrentBib}

\bibitem [\protect \citeauthoryear {%
Waldschl\"ager%
\ \protect \BOthers {.}}{%
Waldschl\"ager%
\ \protect \BOthers {.}}{%
{\protect \APACyear {2022}}%
}]{%
WALDSCHLAGER2022104021}
\APACinsertmetastar {%
WALDSCHLAGER2022104021}%
\begin{APACrefauthors}%
Waldschl\"ager, K.%
, Br\"uckner, M\BPBI Z.%
, {Carney Almroth}, B.%
, Hackney, C\BPBI R.%
, Adyel, T\BPBI M.%
, Alimi, O\BPBI S.%
\BDBL {}Wu, N.%
\end{APACrefauthors}%
\unskip\
\newblock
\APACrefYearMonthDay{2022}{}{}.
\newblock
{\BBOQ}\APACrefatitle {{Learning from natural sediments to tackle microplastics challenges: A multidisciplinary perspective}} {{Learning from natural sediments to tackle microplastics challenges: A multidisciplinary perspective}}.{\BBCQ}
\newblock
\APACjournalVolNumPages{Earth-Science Reviews}{228}{}{104021}.
\PrintBackRefs{\CurrentBib}

\bibitem [\protect \citeauthoryear {%
Waldschl\"ager%
\ \BBA {} Sch\"uttrumpf%
}{%
Waldschl\"ager%
\ \BBA {} Sch\"uttrumpf%
}{%
{\protect \APACyear {2019}}%
{\protect \APACexlab {{\protect \BCnt {1}}}}}]{%
WALDSCHLAGER2019}
\APACinsertmetastar {%
WALDSCHLAGER2019}%
\begin{APACrefauthors}%
Waldschl\"ager, K.%
\BCBT {}\ \BBA {} Sch\"uttrumpf, H.%
\end{APACrefauthors}%
\unskip\
\newblock
\APACrefYearMonthDay{2019{\protect \BCnt {1}}}{}{}.
\newblock
{\BBOQ}\APACrefatitle {Effects of particle properties on the settling and rise velocities of microplastics in freshwater under laboratory conditions} {Effects of particle properties on the settling and rise velocities of microplastics in freshwater under laboratory conditions}.{\BBCQ}
\newblock
\APACjournalVolNumPages{Environmental Science \& Technology}{53}{4}{1958-1966}.
\PrintBackRefs{\CurrentBib}

\bibitem [\protect \citeauthoryear {%
Waldschl\"ager%
\ \BBA {} Sch\"uttrumpf%
}{%
Waldschl\"ager%
\ \BBA {} Sch\"uttrumpf%
}{%
{\protect \APACyear {2019}}%
{\protect \APACexlab {{\protect \BCnt {2}}}}}]{%
Waldschlaeger2019}
\APACinsertmetastar {%
Waldschlaeger2019}%
\begin{APACrefauthors}%
Waldschl\"ager, K.%
\BCBT {}\ \BBA {} Sch\"uttrumpf, H.%
\end{APACrefauthors}%
\unskip\
\newblock
\APACrefYearMonthDay{2019{\protect \BCnt {2}}}{}{}.
\newblock
{\BBOQ}\APACrefatitle {Erosion behavior of different microplastic particles in comparison to natural sediments} {Erosion behavior of different microplastic particles in comparison to natural sediments}.{\BBCQ}
\newblock
\APACjournalVolNumPages{Environ. Sci. Technol.}{53}{22}{13219-13227}.
\PrintBackRefs{\CurrentBib}

\bibitem [\protect \citeauthoryear {%
White%
}{%
White%
}{%
{\protect \APACyear {2016}}%
}]{%
White2016}
\APACinsertmetastar {%
White2016}%
\begin{APACrefauthors}%
White, F\BPBI M.%
\end{APACrefauthors}%
\unskip\
\newblock
\APACrefYear{2016}.
\newblock
\APACrefbtitle {Fluid Mechanics, 8th edition} {Fluid mechanics, 8th edition}\ [Book].
\newblock
\APACaddressPublisher{}{McGraw Hill}.
\PrintBackRefs{\CurrentBib}

\bibitem [\protect \citeauthoryear {%
Wiberg%
\ \BBA {} Smith%
}{%
Wiberg%
\ \BBA {} Smith%
}{%
{\protect \APACyear {1987}}%
}]{%
Wiberg1987}
\APACinsertmetastar {%
Wiberg1987}%
\begin{APACrefauthors}%
Wiberg, P\BPBI L.%
\BCBT {}\ \BBA {} Smith, J\BPBI D.%
\end{APACrefauthors}%
\unskip\
\newblock
\APACrefYearMonthDay{1987}{}{}.
\newblock
{\BBOQ}\APACrefatitle {Calculations of the critical shear stress for motion of uniform and heterogeneous sediments} {Calculations of the critical shear stress for motion of uniform and heterogeneous sediments}.{\BBCQ}
\newblock
\APACjournalVolNumPages{Water Resources Research}{23}{8}{1471-1480}.
\PrintBackRefs{\CurrentBib}

\bibitem [\protect \citeauthoryear {%
Wilcock%
}{%
Wilcock%
}{%
{\protect \APACyear {1988}}%
}]{%
WILCOCK1988}
\APACinsertmetastar {%
WILCOCK1988}%
\begin{APACrefauthors}%
Wilcock, P\BPBI R.%
\end{APACrefauthors}%
\unskip\
\newblock
\APACrefYearMonthDay{1988}{}{}.
\newblock
{\BBOQ}\APACrefatitle {Methods for estimating the critical shear-stress of individual fractions in mixed-size sediment} {Methods for estimating the critical shear-stress of individual fractions in mixed-size sediment}.{\BBCQ}
\newblock
\APACjournalVolNumPages{Water Resour. Res.}{24}{7}{1127-1135}.
\PrintBackRefs{\CurrentBib}

\bibitem [\protect \citeauthoryear {%
Yu%
, Loewen%
, Guo%
, Guo%
\BCBL {}\ \BBA {} Zhang%
}{%
Yu%
\ \protect \BOthers {.}}{%
{\protect \APACyear {2023}}%
}]{%
Yu2023}
\APACinsertmetastar {%
Yu2023}%
\begin{APACrefauthors}%
Yu, Z.%
, Loewen, M.%
, Guo, S.%
, Guo, Z.%
\BCBL {}\ \BBA {} Zhang, W.%
\end{APACrefauthors}%
\unskip\
\newblock
\APACrefYearMonthDay{2023}{}{}.
\newblock
{\BBOQ}\APACrefatitle {Investigation of the Sheltering Effects on the Mobilization of Microplastics in Open-Channel Flow} {Investigation of the sheltering effects on the mobilization of microplastics in open-channel flow}.{\BBCQ}
\newblock
\APACjournalVolNumPages{Environmental Science \& Technology}{57}{30}{11259-11266}.
\PrintBackRefs{\CurrentBib}

\bibitem [\protect \citeauthoryear {%
Yu%
, Yao%
, Loewen%
, Li%
\BCBL {}\ \BBA {} Zhang%
}{%
Yu%
\ \protect \BOthers {.}}{%
{\protect \APACyear {2022}}%
}]{%
Yu2022}
\APACinsertmetastar {%
Yu2022}%
\begin{APACrefauthors}%
Yu, Z.%
, Yao, W.%
, Loewen, M.%
, Li, X.%
\BCBL {}\ \BBA {} Zhang, W.%
\end{APACrefauthors}%
\unskip\
\newblock
\APACrefYearMonthDay{2022}{}{}.
\newblock
{\BBOQ}\APACrefatitle {Incipient Motion of Exposed Microplastics in an Open-Channel Flow} {Incipient motion of exposed microplastics in an open-channel flow}.{\BBCQ}
\newblock
\APACjournalVolNumPages{Environmental Science \& Technology}{56}{20}{14498-14506}.
\PrintBackRefs{\CurrentBib}

\end{thebibliography}

\end{document}